\documentclass[a4paper,aps,pre,twocolumn,notitlepage,intlimits,superscriptaddress,8pt,floatfix]{revtex4-1}
\usepackage[tbtags]{amsmath}
\usepackage{amsthm,amssymb,dsfont,mathrsfs,bbm,xfrac,tabularx}
\usepackage{mathtools}
\usepackage{multirow}
\usepackage[cal=esstix]{mathalfa}
\DeclareMathAlphabet\mathrsfso{U}{rsfso}{m}{n}
\DeclareMathAlphabet\mathzapf{T1}{pzc}{mb}{it}
\usepackage[usenames,dvipsnames]{xcolor}
\usepackage[caption=false]{subfig}
\usepackage[english]{babel}
\usepackage{hyperref}
\hypersetup{pdfauthor={Semerjian Sicuro Zdeborova},pdftitle={Planted matching},%
            colorlinks, linktocpage=true, pdfstartpage=3, pdfstartview=FitV,%
    breaklinks=true, pdfpagemode=UseNone, pageanchor=true, pdfpagemode=UseOutlines,%
    plainpages=false, bookmarksnumbered, bookmarksopen=true, bookmarksopenlevel=1,%
    hypertexnames=true, pdfhighlight=/O,%
    urlcolor=orange, linkcolor=blue, citecolor=red, 
        }
\usepackage{tikz}
\usetikzlibrary{patterns,decorations,arrows,shapes.geometric,arrows,mindmap,backgrounds,positioning,positioning,fit,backgrounds,positioning,arrows,matrix,calc}
\usetikzlibrary{shapes,backgrounds,mindmap,shadows,shapes.geometric,topaths,snakes,arrows,pgfplots.groupplots,fadings,calc,decorations.markings,positioning,chains,fit}

\DeclareMathOperator{\e}{e}
\newcommand{\dd}{{\rm d}}

\DeclareMathOperator{\supp}{supp}

\allowdisplaybreaks
\begin{document}
\title{Recovery thresholds in the sparse planted matching problem}
\author{Guilhem Semerjian}\email{guilhem.semerjian@phys.ens.fr}
\affiliation{Laboratoire de Physique de l'\'Ecole Normale Sup\'erieure, ENS, Universit\'e PSL, CNRS, Sorbonne Universit\'e, Universit\'e de Paris, F-75005 Paris, France}
\author{Gabriele Sicuro}\email{gabriele.sicuro@phys.ens.fr}
\affiliation{Laboratoire de Physique de l'\'Ecole Normale Sup\'erieure, ENS, Universit\'e PSL, CNRS, Sorbonne Universit\'e, Universit\'e de Paris, F-75005 Paris, France}
\author{Lenka Zdeborov\'a}\email{lenka.zdeborova@cea.fr}
\address{Universit\'e Paris-Saclay, CNRS, CEA, Institut de physique th\'eorique, 91191, Gif-sur-Yvette, France}
\date{\today}
\begin{abstract}
We consider the statistical inference problem of recovering an unknown perfect matching, hidden in a weighted random graph, by exploiting the information arising from the use of two different distributions for the weights on the edges inside and outside the planted matching. A recent work has demonstrated the existence of a phase transition, in the large size limit, between a full and a partial recovery phase for a specific form of the weights distribution on fully connected graphs. We generalize and extend this result in two directions: we obtain a criterion for the location of the phase transition for generic weights distributions and possibly sparse graphs, exploiting a technical connection with branching random walk processes, as well as a quantitatively more precise description of the critical regime around the phase transition.
\end{abstract}
\maketitle

\section{Introduction}

A matching of a graph is a subset of its edges such that each node belongs to at most one edge of the matching; in a perfect matching all the nodes are covered in this way~\cite{lovasz2009matching}. In a weighted graph one defines the weight of a matching as the sum of the weights on its edges, and one can try to minimize or maximize this total weight under the (perfect) matching constraint~\cite{Edmonds1965}. This extremization is a problem of combinatorial optimization, widely studied in mathematics, computer science, and also in statistical physics. In this paper we study the planted matching problem, a statistical inference problem where one hides (plants) a perfect matching into a graph, the goal being to find it back. Planted matching problems arise in applications such as particle tracking systems used in experimental physics~\cite{Chertkov2010}; the present paper concentrates on a more fundamental aspect, namely the mathematical description of the peculiar type of phase transition this inference problem exhibits.

Statistical inference on graphs and networks is an area of recent interest including problems such as community detection~\cite{decelle2011asymptotic}, group testing~\cite{mezard2008group}, planted Hamiltonian cycle recovery~\cite{Bagaria2018}, certain types of  error correcting codes~\cite{richardson2008modern}, and many others. All these problems share the common pattern of a signal being observed indirectly via the edges and weights of a graph, with the goal to infer the signal back from these observations. Interestingly as a function of the signal-to-noise ratio and in the limit of large system sizes one encounters sharp thresholds (phase transitions). A classical network-inference problem presenting a phase transition is the community detection in graphs created by the Stochastic Block Model (SBM)~\cite{decelle2011asymptotic}. In the SBM on sparse graphs a phase transition happens between a no-recovery phase where an estimation of the signal better than a random guess is impossible and a partial-recovery phase where a positive (but bounded away from one) correlation with the signal can be obtained. This phase transition can be of second or first order depending on the details of the model, an interesting connection was put forward between phase transitions of first order and existence of algorithmically hard phases~\cite{decelle2011asymptotic}. A typology of phase transitions in problems where the detectability transition (from zero correlation to positive one) appears was recently presented in~\cite{ricci2019typology}. In the SBM, in order to obtain full-recovery of the signal, i.e. a correlation with the signal converging to one, the average degree of the graph has to diverge as the logarithm of its size~\cite{abbe2015exact}. In low-density-parity-check error correcting codes~\cite{richardson2008modern}, another widely studied example of inference problem on sparse graphs, there is also a first order phase transition but this time from the partial-recovery phase to the full-recovery phase where the signal (codeword) can be reconstructed with an error that vanishes as the system size diverges. 

Here we study the planted matching problem, where a perfect matching is hidden in a graph by adding edges to it. The information about which edges were added comes through the distribution of the weights, with different distributions on the planted and non-planted edges. This problem was introduced in~\cite{Chertkov2010} as a toy model in a particle tracking problem, and was studied numerically by solving the corresponding recursive distributional equations for a particular case of the distribution of weights, suggesting a phase transition between full-recovery and partial-recovery phases. More recently, \cite{Moharrami2019} rigorously analysed another special case of the distribution of weights, and proved the existence of such a phase transition. 

We generalize and extend these previous results in several directions (at the level of rigor of theoretical physics). We locate the recovery phase transition for generic weight distributions, considering also a sparse regime for the edges added to the planted configuration. As in~\cite{Chertkov2010} we use the standard cavity method related to the belief propagation approximation~\cite{mezard2009information}. Our key contribution is an analytical insight into how the solution behaves, which allows us to derive a rather simple closed-form expression for the threshold, Eq.~\eqref{condizione_original}, that holds for generic distribution of weights and both sparse and fully connected graphs (the threshold for the sparse case converges to its fully connected limit exponentially fast in the average degree, see for example the phase diagram in Fig.~\ref{fig:phased}). The results of~\cite{Moharrami2019} are also based on the cavity method, but apply only to fully connected graphs and only to the exponential distribution of weights on the planted edges. In this particular case the corresponding recursive distributional equations reduce into a closed system of differential equations. Instead we obtain the generic expression for the threshold by noticing a relation between the solution of the recursive distribution cavity equations and properties of branching random walk processes. The latter, and more generically the phenomenon of front propagation for reaction-diffusion equations, appear in a variety of context and have been extensively studied both in physics and in mathematics~\cite{Hammersley1974,Kingman1975,Biggins1977,Brunet1997,Majumdar2000,Ebert2000,Bachmann2000,Aldous2005,Aidekon2013,Shi15,Bramson2016}; the precise way in which this connection arises here is nevertheless, as far as we know, original in the context of mean-field inference problems. 

Both our work and~\cite{Moharrami2019} show that the recovery threshold in the planted matching problem is of a rather different nature than the thresholds known in the stochastic block model, error correcting codes, or others discussed in the literature. Indeed it separates partial and full recovery phases while occuring at a finite average degree, and there is no sign of a computational gap between the information theoretically optimal reconstruction accuracy and the one achievable by efficient algorithms. Another aspect in which this transition differs from more usual ones is its thermodynamic order: for the specific case studied in~\cite{Moharrami2019} we provide a quantitatively more precise description of the critical regime around the phase transition, Eq.~\eqref{eq_rho_asymptotic}, showing that it is of infinite order in the usual thermodynamic classification (all the derivatives of the order parameter vanish at the transition point). 

The rest of the paper is organized as follows. In Section~\ref{sec:definitions} we define more explicitly the problem under study and introduce two statistical estimators (block and symbol Maximum A Posteriori (MAP)) of the planted matching. In Sec.~\ref{sec:BP} we present the main equations (Belief Propagation and their probabilistic description) that governs the behavior of the problem. In Section~\ref{sec:location} we derive our first main result, namely the location of the phase transition for arbitrary weight distributions (for the block MAP estimator), that is illustrated in Sec.~\ref{sec:examples} on several examples. Our second main result, i.e. a quantitatively more precise description of the critical regime around the phase transition, is explained in Section~\ref{sec:critical}. We then study numerically in Sec.~\ref{sec:symbol_MAP} the threshold of the phase transition for the symbol MAP estimator, while conclusions and perspectives for future work are presented in Sec.~\ref{sec:conclusions}. Some more technical details are deferred to a series of Appendices.

\section{Definitions}\label{sec:definitions}

\subsection{Planted random weighted graphs}

We shall consider weighted graphs denoted $\mathzapf G_0=(\mathzapf V_0, \mathzapf E_0,\underline{w})$, where $\mathzapf V_0 = \{1,\dots,N\}$ is the set of $N$ vertices, $N$ being an even integer, $\mathzapf E_0$ the set of edges (unordered pairs of distinct vertices of $\mathzapf V_0$), and $\underline{w}=\{w_e : e \in \mathzapf E_0\}$ a collection of real weights assigned to each edge of the graph. We endow the set of weighted graphs with a probability distribution, the generation of $\mathzapf G_0$ from this law corresponding to the following steps:
\begin{itemize}
\item One first chooses a perfect matching $\mathzapf{M}$ of $\mathzapf V_0$ uniformly at random among the $(N-1)!!$ possible ones, in other words $\mathzapf{M}$ contains $N/2$ edges, each vertex of $\mathzapf V_0$ belonging to exactly one edge of $\mathzapf{M}$.
\item The edge set $\mathzapf E_0$ of $\mathzapf G_0$ is made of the disjoint union of $\mathzapf{M}$ and additional edges chosen at random: each of the $\binom{N}{2} - (N/2)$ possible edges not already included in $\mathzapf{M}$ is added to $\mathzapf E_0$ with probability $c/N$.
\item The weights $w_e$ are independent random variables, with an absolutely continuous distribution given by the density $\hat{p}$ if $e \in \mathzapf{M}$ and $p$ if  $e \in \mathzapf E_0 \setminus \mathzapf{M}$.
\end{itemize}

We shall call planted (resp.~non-planted) edges those in $\mathzapf{M}$ (resp.~in $\mathzapf E_0 \setminus \mathzapf{M}$). The parameters of this random ensemble of weighted graphs are thus the even integer $N$, the parameter $c$ controlling the density of non-planted edges, and the two distributions $\hat{p}$ and $p$ for the generation of the weights of the planted and non-planted edges. In formula the probability to generate a graph $\mathzapf G_0$, given the choice of $\mathzapf{M}$, translates from the above description as:
\begin{multline} \label{eq_proba_direct}
\mathbb P(\mathzapf G_0 | \mathzapf M)= \prod_{e \in \mathzapf M} \hat p(w_e) \prod_{e \in \mathzapf E_0 \setminus \mathzapf M} p(w_e) \\ \times\left(\frac{c}{N} \right)^{|\mathzapf E_0|-\frac{N}{2}} \left(1-\frac{c}{N} \right)^{\binom{N}{2} - |\mathzapf E_0|} \mathbb{I}(\mathzapf M \subseteq \mathzapf E_0) \ ,
\end{multline}
where here and in the following $\mathbb{I}(A)$ denotes the indicator function of the event $A$. Note that the number of non-planted edges concentrate in the large size (thermodynamic) limit $N \to \infty$ around its average value $cN/2$, assuming $c$ remains fixed in this limit, and that these edges form essentially an Erd\H os-R\'enyi random graph of average degree $c$ (modulo the exclusion of the planted edges).

The model studied in~\cite{Chertkov2010,Moharrami2019} corresponds to a dense, or fully-connected, version of the model defined above, in which $\mathzapf G_0$ is a complete weighted graph, $\mathzapf E_0$ containing all the possible edges between the $N$ vertices. In order to have a well-defined thermodynamic limit in this dense case it is necessary to rescale with $N$ the weights on the non-planted edges, i.e. to use an $N$-dependent distribution $p$. The simplest way to perform this rescaling is to use $p(w)=Q(w/N)/N$, where $Q$ is a density with a support included in the non-negative reals, and a positive density $Q(0)>0$ at the origin~\cite{Mezard1985,Chertkov2010,Moharrami2019}; without loss of generality we assume $Q(0)=1$. The thermodynamic limit of this dense model is then equivalent to the large degree limit of the sparse one, $c \to \infty$ after $N\to\infty$, if one uses for the distribution $p$ of the non-planted edges the uniform distribution on the interval $[0,c]$. We shall thus study the richer sparse model, with finite $c$, and take the large degree-limit when needed in order to compare our results with those of the dense case.

\subsection{A statistical inference problem}
\label{sec_def_inference}

The question we shall investigate in the following is whether the observation of a graph $\mathzapf G_0$ generated according to the procedure above allows to infer the hidden matching $\mathzapf M$, assuming the observer knows the parameters $c$, $p$ and $\hat p$ of the model. In this setting all the information the observer can exploit to perform this task is contained in the posterior probability $\mathbb P(\mathzapf M|\mathzapf G_0)$. From the expression \eqref{eq_proba_direct} of the graph generation probability, and the knowledge that the prior probability on $\mathzapf M$ is uniform over the set of all perfect matchings, Bayes theorem yields immediately the following expression for the posterior:
\begin{equation}
\mathbb P(\mathzapf M | \mathzapf G_0 ) \propto \prod_{e \in \mathzapf M} \hat p(w_e) \prod_{e \in \mathzapf E_0 \setminus \mathzapf M} p(w_e) \mathbb{I}(\mathzapf M \subseteq  \mathzapf E_0 ) \mathbb{I}_{\rm pm} (\mathzapf M) \ ,\nonumber
\end{equation}
where the symbol $\propto$ hides a normalization constant independent of $\mathzapf M$, and the last term is the indicator function of the event ``$\mathzapf M$ is a perfect matching''. For notational simplicity it is convenient to encode a set of edges $\mathzapf M \subseteq  \mathzapf E_0$ with binary variables, 
$\underline{m}=\{m_e : e \in \mathzapf E_0\} \in \{0,1\}^{\mathzapf E_0}$, where $m_e=1$ if and only $e \in \mathzapf M$, and rewrite the posterior as
\begin{equation}\label{post}
\mathbb P( \underline{m}| \mathzapf G_0 ) \propto \prod_{e \in \mathzapf E_0 } \left( \frac{\hat p(w_e)}{p(w_e)}  \right)^{m_e}  \prod_{i=1}^N \mathbb{I}\left(\sum_{e \in \partial i} m_e = 1 \right) \ ,
\end{equation}
where $\partial i$ denotes the set of edges incident to the vertex $i$. The observer will now compute an estimator $\widehat{\mathzapf M}(\mathzapf G_0)$, this function of the observations being ``as close as possible'' to the hidden matching $\mathzapf M$. The optimal estimator actually depends on which notion of ``closeness'' between $\mathzapf M$ and the estimator $\widehat{\mathzapf M}(\mathzapf G_0)$ is used. 

If the measure of the distance between them is simply the indicator function $\mathbb{I}( \mathzapf M \neq \widehat{\mathzapf M}(\mathzapf G_0))$, then the optimal estimator, optimal in the sense that it minimizes this distance averaged over all realizations of the problem, is the one maximizing the posterior,
\begin{equation}\label{eq_bMAP}
\widehat{\mathzapf M}_{\rm b}(\mathzapf G_0) = \underset{\underline{m}}{\rm argmax}\ \mathbb P( \underline{m}| \mathzapf G_0 ) \ ,
\end{equation}
where we slightly abused notations and used freely the equivalence between $\underline{m}$ and $\mathzapf M$. Following the nomenclature of error-correcting codes \cite{richardson2008modern} we shall call this the block Maximal A Posteriori (MAP) estimator, hence the subscript $\rm b$. As we shall detail below the estimator $\widehat{\mathzapf M}_{\rm b}$ is the perfect matching of $\mathzapf G_0$ which minimizes the sum of some effective weights on the edges it contains.

If instead the distance to be minimized is the total number of misclassified edges, $|\mathzapf M \triangle \widehat{\mathzapf M}(\mathzapf G_0)|$, with $\triangle$ the symmetric difference between sets, or equivalently the Hamming distance between the binary strings $\underline{m}$ and $\widehat{\underline{m}}(\mathzapf G_0)$ encoding them, then the optimal estimator is the so-called symbol MAP one, denoted $\widehat{\mathzapf M}_{\rm s}(\mathzapf G_0) $, defined by the binary string $\widehat{\underline{m}}(\mathzapf G_0)$ where, for all the edges $e \in \mathzapf E_0$,
\begin{equation} \label{eq_sMAP}
\widehat{m}_e(\mathzapf G_0) = \underset{m_e}{\rm argmax}\ \mathbb P_e(m_e | \mathzapf G_0 ) \ ,
\end{equation}
with $\mathbb P_e$ the marginal of the posterior probability \eqref{post} for the edge $e$. Note that this estimator is not necessarily a perfect matching, nevertheless it is the one that minimizes the distance $|\mathzapf M \triangle \widehat{\mathzapf M}(\mathzapf G_0)|$ on average over all realizations of the problem.

For future use let us define the (reduced) distance between the planted matching $\mathzapf M$ and an arbitrary estimator $\widehat{\mathzapf M}$ (i.e. a subset of the edge set $\mathzapf E_0$) as
\begin{align}
& \varrho(\mathzapf M,\widehat{\mathzapf M}) = \frac{1}{N}|\mathzapf M \triangle \widehat{\mathzapf M}| \\ &= \frac{1}{N} \sum_{e \in \mathzapf M } \mathbb{I}(\widehat{m}_e=0) + \frac{1}{N} \sum_{e \in \mathzapf E_0 \setminus \mathzapf M } \mathbb{I}(\widehat{m}_e=1) \ . 
\label{eq_def_rho}
\end{align}
If $\widehat{\mathzapf M}$ contains exactly $N/2$ edges, in particular if it is a perfect matching, this expression can be simplified as the two terms contribute in the same way (there are as many false positive as true negative errors in the identification of the edges), then
\begin{equation}\label{eq_rho_simplified}
\varrho(\mathzapf M,\widehat{\mathzapf M}) =\frac{2}{N} \sum_{e \in \mathzapf M } \mathbb{I}(\widehat{m}_e=0) \ .
\end{equation}

Our goal in the rest of the article is to discuss the quality of the estimators defined above, in the thermodynamic limit $N \to \infty$, as a function of the parameters of the model. Following the studies of~\cite{Chertkov2010,Moharrami2019} one expects to find phase transitions between full recovery phases, in which all but a vanishing fraction of the edges of $\mathzapf M$ can be recovered from the observation of $\mathzapf G_0$, characterized by a vanishing average reconstruction error $\mathbb{E}[\varrho]=0$, and partial recovery phases where a positive fraction of the edges will be misclassified, $\mathbb{E}[\varrho]>0$. Before entering the actual computations let us make two simple remarks in order to give the reader a first intuitive idea of the effect of the parameters on the inference difficulty. (i) The identification of the planted edges will be easier if the distributions $p$ and $\hat{p}$ are less similar one to the other; in the extreme cases where $p=\hat{p}$ the weights contain absolutely no information on $\mathzapf M$, while if $p$ and $\hat{p}$ have disjoint supports $\mathzapf M$ can be identified by a simple inspection of the weights on the edges. (ii) For a fixed choice of $p$ and $\hat p$ the parameter $c$ corresponds to a noise level: if $c$ is very small $\mathzapf E_0$ contains essentially only the sought-for edges of~$\mathzapf M$, increasing it the latter are hidden in the confusing non-planted edges.

\section{Cavity method equations}\label{sec:BP}

\subsection{A first pruning of the graph}\label{sec:pruning}
Before proceeding further, let us observe that the inference problem can be in general reduced in size after some simple, preliminary observations. Following the remark (i) in a less drastic case, suppose that the supports of $p$ and $\hat p$ are different (but not necessarily disjoint). Then an edge $e$ bearing a weight $w_e$ in the support of $p$ but not in the one of $\hat p$ is, without doubt, non-planted; conversely $e$ is certainly planted if $w_e$ is in the support of $\hat p$ but not in the one of $p$. All the edges identified in this way can be eliminated from $\mathzapf G_0$; moreover the two vertices belonging to an edge identified as planted can also be eliminated, as well as the other edges incident to them, that cannot be planted by definition of a perfect matching.

To put these remarks on a quantitative ground let us denote $\supp(\hat p)\coloneqq\{w\in\mathds{R}\colon \hat p(w)>0\}$ (more precisely the closure of this set) the support of the distribution $\hat p$, and similarly $\supp(p)\coloneqq\{w\in\mathds{R}\colon p(w)>0\}$ the support of $p$. We define
\begin{subequations}
\begin{align}
\Gamma&\coloneqq\supp(p)\cap\supp(\hat p) \ , \label{def:gamma}\\
\mu&\coloneqq\int_{\Gamma}p(w)\dd w \ , \\
\hat\mu&\coloneqq\int_{\Gamma}\hat p(w)\dd w.
\end{align}
\end{subequations}
A non-planted edge $e$ has weight $w_e\not\in\Gamma$, and can thus be identified, with probability $1-\mu$. Similarly, a planted edge $e$ will have $w_e\not\in\Gamma$ with probability $1-\hat\mu$, and, in this case, it is surely an element of $\mathzapf M$. We will denote
\begin{equation}
{\mathzapf M}_0\coloneqq \left\{e\in\mathzapf E_0\colon w_e \in \supp(\hat p) \setminus \supp(p) \right\}
\end{equation}
the set of planted edges immediately recognizable by means of these simple considerations. The edges in ${\mathzapf M}_0$ can be removed from the graph, alongside with their endpoints and all edges incident to them. After this pruning process, the obtained graph has, on average and in the large $N$ limit, $\hat\mu N$ surviving vertices, each of them with degree $1+\mathsf Z$, $\mathsf Z$ being a Poisson random variable of mean $\gamma\coloneqq c\mu\hat\mu$ (each non-planted edge is present with probability $\mu\frac{c}{N}$, but in a graph with $\hat\mu N$ vertices).

The distribution of the weights of the surviving edges is now conditioned to the fact that their values are in $\Gamma$. On the new pruned graph, therefore, the weight distributions are
\begin{subequations}\label{PPhat}
\begin{align}
P(w)&\coloneqq \frac{1}{\mu}p(w)\mathbb I(w\in\Gamma) \ , \\
\hat P(w) &\coloneqq \frac{1}{\hat\mu}\hat p( w)\mathbb I( w\in\Gamma) \ ,
\end{align}
\end{subequations}
for the non-planted and planted edges, respectively.
We will denote $\mathzapf G =(\mathzapf V, \mathzapf E, \underline{w})$ the graph obtained after this pruning, with $\mathzapf V\subseteq\mathzapf V_0$ and $\mathzapf E \subseteq \mathzapf E_0$ the new vertex and edge sets.

\subsection{The Belief Propagation equations}

Here we present the belief propagation algorithm that was used in \cite{Chertkov2010} for the planted matching problem. Let us introduce a positive parameter (fictitious inverse temperature) $\beta$, and consider the following probability distribution over the configurations $\underline{m}=\{m_e : e \in \mathzapf E\} \in \{0,1\}^{\mathzapf E}$ of binary variables on the edges of a weighted graph $\mathzapf G$,
\begin{equation}\label{lbeta}
\nu(\underline{m})\propto \e^{-\beta\sum_{e \in\mathzapf E} m_e \omega_e} \prod_{i \in\mathzapf V}\mathbb I\left(\sum_{e\in \partial i}m_e=1\right) \ ,
\end{equation}
where we introduced effective weights $\omega_e$ on the edges, that are computed from the observed weights $w_e$ as $\omega_e = \omega(w_e)$, with
\begin{equation}\label{omega} 
\omega(w)\coloneqq -\ln\frac{\hat P(w)}{P(w)} \ .
\end{equation}
To lighten the notation we kept implicit the dependency on $\beta$ and $\mathzapf G$ of the probability distribution $\nu$; for $\beta=1$ it coincides with the posterior defined in \eqref{post}, when $\beta \to \infty$ it concentrates on the configurations maximizing the posterior, these two values of $\beta$ allow thus to deal with the symbol and block MAP estimators, respectively.

The probability distribution $\nu$ defined in Eq.~\eqref{lbeta} has the form of a Gibbs measure over all weighted perfect matchings of the graph $\mathzapf G$. The exact computation of its marginals is an intractable task in general; we shall instead study it in an approximate way, using the Belief Propagation (BP) algorithm (see for instance~\cite{mezard2009information} for a general introduction to BP as well as chapter 16 therein for its application to matching problems), that is conjectured to provide an asymptotically exact description in the large size limit for these sparse random graphs. One can indeed consider Eq.~\eqref{lbeta} as a graphical model, with variable nodes $m_e$ living on the edges of $\mathzapf G$, and two types of interaction nodes: one on each vertex $i \in \mathzapf V$, that imposes that exactly one variable $m_e$ is equal to 1 around it, and one ``local field'' interaction $\e^{-\beta m_e \omega_e} $ for each variable. The BP equations are then obtained by introducing ``messages'' on the edges of this factor graph, that mimic the marginal probabilities in amputated graphical models and would become exact if the factor graph were a tree. For the model at hand these messages are of the form $\nu_{i\to e}(m)$, from a vertex $i$ to an edge $e=(i,j)$, and obey the following equations (one for each directed edge of the graph),
\begin{multline}
 \nu_{i\to e}(m)\propto\\\quad
 \sum_{\mathclap{\{m_{\tilde e}\}_{\tilde e\in\partial i\setminus e}}}\ \mathbb I\left(\!m\!+\!\sum_{\mathclap{\tilde e\in\partial i\setminus e}}m_{\tilde e}=1\!\right)\prod_{\mathclap{\substack{\tilde e=(r,i)\\\tilde e\in\partial i\setminus e}}}\nu_{r\to \tilde e}(m_{\tilde e})\e^{-\beta m_{\tilde e}\omega_{\tilde e}} \ .
\end{multline}
We adopt the convention $\sum_{a\in A}f(a)=0$ and $\prod_{a\in A}f(a)=1$ if $A=\emptyset$ for any function $f$. As the variables are binary, $m_e\in\{0,1\}$ for each $e\in\mathzapf E$, the messages can be conveniently parametrized in terms of ``cavity fields'' $h_{i\to e}$, one real number for each directed edge, as
\begin{equation}
 \nu_{i\to e}(m)\coloneqq\frac{\e^{\beta m h_{i\to e}}}{1+\e^{\beta h_{i\to e}}}\ ,
\end{equation}
so that the BP equations become in terms of the cavity fields:
\begin{equation}\label{cavbeta}
h_{i\to e}=-\frac{1}{\beta}\ln\left[\sum_{\tilde e=(r,i)\in\partial i\setminus e}\mkern-20mu\e^{-\beta(\omega_{\tilde e} - h_{r\to \tilde e})}\right] \ .
\end{equation}
Once a solution of the set of BP equations has been found (for instance by iterating them starting from a random or zero initial condition until convergence to a fixed point is reached), the BP approximation of the marginal probability of the variable $m_{e}$ on the edge $e=(i,j)$ is given by
\begin{equation}\label{marginal}
 \nu_{e}(m)=\frac{\e^{\beta m(h_{i\to e}+h_{j\to e}-\omega_{e})}}{1+\e^{\beta (h_{i\to e}+h_{j\to e}-\omega_{e})}} \ .
\end{equation}

The BP approximation to the symbol MAP estimator defined in \eqref{eq_sMAP} is thus obtained by solving the BP equations with $\beta=1$, and estimating as a planted edge those for which $\nu_{e}(1)>1/2$, namely
\begin{equation}
\begin{split}
\widehat{\mathzapf M}_{\rm s}(\mathzapf G)&\coloneqq \left\{e\in\mathzapf E\colon \nu_e(1)>\frac{1}{2}\right\}\\
&=\left\{e=(i,j)\in\mathzapf E\colon h_{i\to e}+h_{j\to e} > \omega_{e}\right\} \ .
\end{split}
\label{eq_inclusion_rule}
\end{equation}

We will keep the same rule (\ref{eq_inclusion_rule}) for the conversion of a solution of the BP equations into an estimator of $\mathzapf M$ for all values of $\beta$, and in particular for $\beta\to\infty$. If the block MAP configuration is unique, and if the marginal probabilities $\nu_e$ are computed exactly, then this is a legitimate way of determining the block MAP estimator \eqref{eq_bMAP}. The BP algorithm is of course only an approximation here, but we conjecture it to be asymptotically exact, i.e. that the reduced Hamming distance between the block MAP estimator and its BP version vanishes in the thermodynamic limit. This relies on rigorous works that, even if they do not directly apply to the case considered here, have proven the exactness of the BP algorithm in similar settings. More precisely, \cite{Bayati2008} proved that for a given bipartite weighted graph, if the perfect matching with minimal weight is unique then the $\beta\to\infty$ version of the BP equations, associated to the inclusion rule \eqref{eq_inclusion_rule}, converges to the optimal configuration, in a number of iterations that scale with the gap between the optimal weight and the second minimum, and with the largest weight in the graph. \cite{Salez2009} improved this convergence rate for typical bipartite graphs of a random ensemble, while \cite{Sanghavi2011,Bayati2011} removed the bipartiteness assumption but added an hypothesis on the absence of fractional solutions for the Linear Programming relaxation of the problem.

\subsection{Recursive Distributional Equations}

The BP equations have been written in Eqs.~\eqref{cavbeta} for a given instance of the graph $\mathzapf G$; to obtain the average error on the ensemble of all possible instances of our problem we need to describe the statistics of the solutions of the BP equations. This step is known as density evolution in the context of error-correcting codes, or as the cavity method in statistical mechanics~\cite{Mezard2001}. We refer the reader to~\cite{Zdeborova2006a,Bordenave2013} for similar studies of the matchings in sparse, non-planted random graphs, and to~\cite{Gamarnik2006,Parisi2020} which considered the weighted case (still without a planted structure).

Suppose that an instance is generated at random, that the BP equations are solved on it, and that a directed planted edge is chosen uniformly at random, say $i \to e$; let us call $\hat {\mathsf H}$ the random variable that has the law of the cavity field $h_{i \to e}$. We define similarly ${\mathsf H}$ as the random variable distributed as $h_{i \to e}$ when one chooses a non-planted edge. Let us also introduce the random variables $\Omega=\omega(W)$ and $\hat\Omega=\omega(\hat W)$, where $W$ (resp. $\hat W$) is a random variable with density $P$ (resp. $\hat P$). If one assumes that the typical realizations of $\nu$ have no long-range correlations (the so-called replica symmetric (RS) hypothesis), then \eqref{cavbeta} translates into  recursive distributional equations (RDEs) between the random variables ${\mathsf H}$ and $\hat {\mathsf H}$. A vertex $i$ in a directed planted edge $i \to e$ is incident to a Poissonian number of other non-planted edges because of the Erd\H os-R\'enyi nature of the latter, and similarly if $i \to e$ is non-planted there will be exactly one planted edge incident to $i$, and other non-planted edges from the Erd\H os-R\'enyi part of the graph. With the RS assumption of independence of the incoming cavity fields one thus obtains:
\begin{subequations}\label{rdebeta}
\begin{align}
\hat {\mathsf H}&\stackrel{\mathrm d}{=}-\frac{1}{\beta}\ln\left(\sum_{i=1}^{\mathsf Z}\e^{-\beta(\Omega_i-{\mathsf H}_i)}\right)\ , \label{cavbetaH1}\\
{\mathsf H}&\stackrel{\mathrm d}{=}-\frac{1}{\beta}\ln\left(\e^{-\beta(\hat \Omega-\hat {\mathsf H})} +\sum_{i=1}^{\mathsf Z}\e^{-\beta(\Omega_i-{\mathsf H}_i)}\right) \nonumber \\
&\stackrel{\mathrm d}{=}-\frac{1}{\beta}\ln\left(\e^{-\beta(\hat \Omega-\hat {\mathsf H})} + \e^{-\beta\hat {\mathsf H}'}\right)\ .\label{cavbetaH2}
\end{align}
\end{subequations}
In the equations above all random variables are independent, $\mathsf Z$ is Poisson distributed with mean $\gamma$, the $\Omega_i$'s have the same law as $\Omega$, and similarly ${\mathsf H}_i$ are independent copies of $\mathsf H$, and $\hat {\mathsf H}'$ of $\hat {\mathsf H}$.

The average of the reconstruction error defined in \eqref{eq_def_rho} can be computed in this setting, recalling the inclusion rule \eqref{eq_inclusion_rule}, as:
\begin{equation}
\mathbb{E}[\varrho] = \frac{\hat \mu}{2} \mathbb{P}[\hat {\mathsf H} + \hat {\mathsf H}' \le \hat \Omega  ]
+ \frac{\hat \mu \gamma }{2} \mathbb{P}[\mathsf H + \mathsf H ' > \Omega  ] \ .
\label{eq_averho_1}
\end{equation}

\subsection{A second pruning of the graph}
\label{sec_second_pruning}

Our goal in the following will be to understand the properties of the random variables ${\mathsf H}$ and $\hat {\mathsf H}$ solutions of \eqref{rdebeta}, and their possible bifurcations when the parameters of the model are varied. As a first step in this direction we shall isolate the contribution of ``hard-fields'', in other words the probabilities of the events $\hat {\mathsf H} = + \infty$ and ${\mathsf H} = - \infty$ for these random variables. Observe indeed that $\mathbb{P}[\mathsf Z=0]>0$ in \eqref{cavbetaH1}, which leads to $\hat {\mathsf H} = + \infty$, and that this event implies ${\mathsf H} = - \infty$ in \eqref{cavbetaH2}.  From both theoretical and practical point of views it is convenient to deal with these events explicitly, we shall thus introduce the probabilities $\hat q$ and $q$ of  the events $\hat {\mathsf H} = + \infty$ and ${\mathsf H} = - \infty$ respectively, and two new random variables $\hat H$ and $H$ that have the law of $\hat {\mathsf H}$ and ${\mathsf H}$ conditional on being finite (we exclude the possibility of $\hat {\mathsf H} = - \infty$ and ${\mathsf H} = + \infty$,  $\hat H$ and $H$ are finite with probability one).
In formulas these definitions amount to
\begin{subequations}
\begin{align}
\mathsf H & \stackrel{\mathrm d}{=}\begin{cases} -\infty&\text{with prob. }q \ ,\\
           H&\text{with prob. } 1-q \ ,
          \end{cases}
\\
\hat{\mathsf H} & \stackrel{\mathrm d}{=}\begin{cases} +\infty&\text{with prob. }\hat q \ ,\\
           \hat H&\text{with prob. } 1-\hat q \ .
          \end{cases}
\end{align}
\end{subequations}
Let us insert them in \eqref{rdebeta} in order to obtain the equations obeyed by $q$, $\hat q$, $H$ and $\hat H$. In the right hand side of \eqref{cavbetaH1} the number of infinite and finite $\mathsf H_i$'s are easily seen to be two independent Poisson random variables of parameters $\gamma q$ and $\gamma (1-q)$ respectively. $\hat{\mathsf H}$ is infinite if and only if the second of this number vanishes, hence one has $\hat q = \e^{-\gamma(1-q)}$ and
\begin{equation}
\hat {H}\stackrel{\mathrm d}{=}-\frac{1}{\beta}\ln\left(\sum_{i=1}^{Z}\e^{-\beta(\Omega_i- {H}_i)}\right) \ ,
\end{equation}
where $Z$ has the law of a Poisson random variable of parameter $\gamma (1-q)$ conditioned to be strictly positive, i.e. $\mathbb{P}[Z=k] =\frac{(\gamma (1-k))^k}{k! (\e^{\gamma(1-q)}-1 ) } $ for $k \ge 1$.
In \eqref{cavbetaH2} one sees that $\mathsf H$ is infinite if and only if $\hat{\mathsf H}$ is infinite, hence $q=\hat q$ and
\begin{equation}
H \stackrel{\mathrm d}{=}
\begin{cases}
\hat \Omega-\hat{H}&\text{with prob. } \hat q \ , \\
-\frac{1}{\beta}\ln\left( \e^{-\beta (\hat \Omega - \hat {H})} + \e^{-\beta\hat {H}'}\right)
& \text{with prob. }1- \hat q \ .
\end{cases}
\end{equation}

In summary the elimination of the hard fields amount to find the solution $q$ of
\begin{equation}
q = \e^{-\gamma(1-q)} \ ,
\label{eq_q}
\end{equation}
and to study the finite random variables $H$ and $\hat H$ solution of the RDEs
\begin{subequations}\label{rdebetaf}
\begin{align}
\hat {H}&\stackrel{\mathrm d}{=}-\frac{1}{\beta}\ln\left(\sum_{i=1}^{Z}\e^{-\beta(\Omega_i- {H}_i)}\right) \ ,
\label{cavbetaH1f}\\
{H}&\stackrel{\mathrm d}{=}
\begin{cases}
\hat \Omega-\hat{H}&\text{with prob. }q \ ,\\
-\frac{1}{\beta}\ln\left(\e^{-\beta (\hat \Omega - \hat {H})} + \e^{-\beta\hat {H}'}\right)
&\text{with prob. }1-q \ ,
\end{cases}\label{cavbetaH2f}
\end{align}
\end{subequations}
where the variable $Z$ in Eq.~\eqref{cavbetaH1f} has distribution $\mathbb{P}[Z=k] = \pi_k$ with 
\begin{equation}\label{disZhat}
\pi_k\coloneqq \frac{q}{1-q}\frac{[(1-q)\gamma]^k}{k!} \qquad \text{for } k \ge 1 \ .
\end{equation}
The average reconstruction error \eqref{eq_averho_1} can be reexpressed in terms of the new random variables $H$ and $\hat H$ as:
\begin{align}
\mathbb{E}[\varrho] =& \frac{\hat \mu (1-q)^2}{2} \mathbb{P}[\hat  H + \hat H' \le \hat \Omega  ] 
\nonumber \\
& + \frac{\hat \mu (1-q)^2\gamma }{2} \mathbb{P}[H + H ' > \Omega  ] \ .
\label{eq_averho_2}
\end{align}

This procedure of ``hard-fields'' elimination that we explained on the RDE's admits also an interpretation on a single graph instance. As a matter of fact the presence of infinite fields on the planted edges can be traced back to the BP equation \eqref{cavbeta} which shows that $h_{i \to e}= +\infty$ if $i$ is a leaf of the graph, i.e. $i$ is of degree 1 and its only incident edge is $e$. But this fact allows us to unambiguously identify $e$ as an edge of the planted matching, which by definition covers all the vertices of the graph. Then $i$, the edge $e=(i,j)$, the vertex $j$ and all the edges incident to it can be removed, the latter being with certainty non-planted edges. This leaf removal procedure can be iterated until either all the graph has been pruned, or stops when a non-trivial core without any leaf has been reached. The propagation of infinite fields in the BP equations is an equivalent way of describing this pruning algorithm. Note that such a leaf removal procedure has already been studied in the literature for standard Erd\H os-R\'enyi graphs \cite{Karp1981,Bauer2001}, in this case a core percolation transition is found when the average degree of the graph crosses the Euler number value $e$: for sparser graphs the leaf removal procedure typically destroys the whole graph, while a non-trivial core survives for larger average degrees. In our case the core percolation transition happens at $\gamma=1$ (which corresponds to the usual percolation transition of the Erd\H os-R\'enyi random graph superposed to the planted matching): if $0<\gamma\leq 1$ the self-consistent equation \eqref{eq_q} on $q$ only admits the solution $q=1$, which means that the leaf removal procedure allows to recover completely the planted matching (up to a subextensive number of edges in the thermodynamic limit). On the contrary for $\gamma > 1$ the leaf removal stops with a non-trivial core  (this explains why one should take the solution $q<1$ of \eqref{eq_q} when $\gamma>1$). We shall see in the following that full recovery phases can exist also for $\gamma > 1$, but in that case the simple leaf removal procedure is not able to identify all the edges of the planted matching, the full recovery is due to a non-trivial amplification effect, by the iterations of the BP equations, of the information contained in the weights of the edges of the core.

\section{The location of the phase transition for the block MAP estimator}
\label{sec:location}

\subsection{RDEs for the block MAP}

As explained in Sec.~\ref{sec_def_inference} the block MAP estimator, that maximizes the probability of correct identification of the whole planted matching, is the configuration $\underline{m}$ that maximizes the posterior in Eq.~\eqref{post}, and can be obtained by taking the ``zero-temperature'' limit $\beta \to \infty$ in the probability distribution $\nu$ defined in \eqref{lbeta}. The BP equations that we wrote in \eqref{cavbeta} for a generic value of $\beta$ become in this limit
\begin{equation}\label{cav1}
h_{i\to e}=\min_{{\tilde e=(r,i)\in \partial i\setminus e}}\left[\omega_{\tilde e}-h_{r\to \tilde e}\right] \ ;
\end{equation}
the configuration maximizing the posterior \eqref{post} can be equivalently defined as the perfect matching of minimum cost on the weighted graph $(\mathzapf V, \mathzapf E,\underline{\omega})$, with the effective weights $\omega_e$ replacing the observed weights $w_e$. Hence these BP equations coincide with those written in \cite{Bayati2008} to study such minimum weight matching problems. With the inclusion criterion of \eqref{eq_inclusion_rule} the BP approximation for the block MAP configuration is determined as
\begin{equation}\label{mij}
 m_{e}=\mathbb{I}(h_{i\to e}+h_{j\to e}-\omega_{e} > 0) \ .
\end{equation}

The probabilistic treatment of the BP equations can also be specialized very easily to the limit case $\beta\to+\infty$, in particular the RDEs \eqref{rdebetaf} yield
\begin{subequations}\label{cavmapmatch}
\thinmuskip=0mu
 \begin{align}
  \hat H&\stackrel{\mathrm{d}}{=}\min_{1\leq i\leq Z}\left[\Omega_{i}-H_{i}\right]\ , \label{cavmapmatch1}\\
  H&\stackrel{\mathrm{d}}{=}
  \begin{cases}
  \hat \Omega-\hat H&\text{ with prob. $q$} \ ,\\
  \min\left(\hat \Omega-\hat H , \ \hat H'\right)&\text{ with prob. $1-q$}\ ,
  \end{cases}  \label{cavmapmatch2}
 \end{align}
\end{subequations}
with the law of the random variable $Z$ defined in \eqref{disZhat}. The average reconstruction error \eqref{eq_averho_2} can actually be simplified for this $\beta \to \infty$ situation into
\begin{equation}\label{errore}
\mathbb{E}[\varrho]=\hat\mu(1-q)^2\mathbb{P}[\hat H +\hat H' \le \hat\Omega] \ ;
\end{equation}
as discussed in Sec.~\ref{sec_def_inference} (see in particular \eqref{eq_rho_simplified}) there are as many misclassified planted and non-planted edges when the estimator contains $N/2$ edges, which we argued to be asymptotically the case for $\beta\to\infty$, hence the two terms in \eqref{eq_averho_2} are equal. For completeness we show in Appendix~\ref{app_rho_blockMAP} that this equality follows indeed from the RDE \eqref{cavmapmatch}, modulo an hypothesis of continuity for the distributions of $H$ and $\hat H$, that mimics the hypothesis of uniqueness of the block MAP assignment.

The equalities in distribution between random variables stated in \eqref{cavmapmatch} can be equivalently rephrased as equations between the cumulative distribution functions of the variables $H$ and $\hat H$. For a random variable $X$ we shall define the c.d.f. $F_X$ and its reciprocal $\bar{F}_X$ according to
\begin{equation}
F_X(x)=\mathbb{P}[X \le x]\ ,\quad \bar{F}_X(x)=1-F_X(x)=\mathbb{P}[X > x] \ .
\nonumber
\end{equation}
One obtains from \eqref{cavmapmatch} 
\begin{subequations}\label{AAhat}
\begin{multline}
\bar{F}_H(h)=\\
\medmuskip=0mu
\thinmuskip=0mu
\thickmuskip=0mu
=(q+(1-q)\bar{F}_{\hat H}(h))\int_{\Gamma} F_{\hat H}(\omega( w)-h) \hat P(w) \dd w\ , \end{multline}
and                                                                                            
\begin{multline}
\medmuskip=0mu
\thinmuskip=0mu
\thickmuskip=0mu
\bar{F}_{\hat H}(h)
=\frac{q}{1-q}\sum_{k=1}^\infty\frac{[\gamma(1-q)]^k}{k!}\left(\int_\Gamma F_H(\omega(w)-h)P(w)\dd w\right)^k\\
=\frac{\exp\left[-\gamma(1-q)\int_\Gamma\bar{F}_H (\omega(w)-h)P(w)\dd w\right]-q}{1-q}.
\end{multline}
\end{subequations}
These are integral non-linear equations on the two functions $F_H$ and $F_{\hat H}$, describing the thermodynamic limit of the planted matching problem. These recursive equations can also be understood as describing the optimal matching problem on an infinite tree. The authors of ~\cite{Moharrami2019} show rigorously that for fully connected graphs the optimal configuration of the finite graph locally converges to the optimal matching of the infinite tree. 

The above integral non-linear equations are quite complicated to solve in general. It was shown in~\cite{Moharrami2019}, that, for a rather specific case (in the large degree limit with an exponential distribution for the planted weights) these integral equations can be transformed into a system of ordinary differential equations (which we will detail in Section \ref{sec:critical}). Unfortunately such a simplification does not seem to hold besides this special case.

The question now is to understand the solution of \eqref{cavmapmatch} (or equivalently of \eqref{AAhat}) as a function of the parameters of the model. It is easy to check that $\hat H = +\infty$, $H= -\infty$ (i.e. $F_{\hat H}(h)=0$, $F_H(h)=1$) is always a solution, for every choice of the parameters. If this is the correct solution then $\mathbb{E}[\varrho]=0$, in other words one is in a full recovery phase. The picture that emerges from the previous works~\cite{Chertkov2010,Moharrami2019} is that for some value of the parameters another solution of \eqref{cavmapmatch} exists, and is attractive when running BP from an initial condition uncorrelated with the planted matching. One is then in a partial recovery phase, with $\mathbb{E}[\varrho] > 0$. On the contrary if $\hat H = +\infty$, $H = -\infty$ is the only solution of \eqref{cavmapmatch} one is in a full recovery phase, the hidden matching being sufficiently attractive to drive the iterations of the BP equations towards it. According to this description the phase transition between full and partial recovery corresponds to the disappearance of a non-trivial solution of the RDE (here and in the following non-trivial means distinct from $\hat H = +\infty$, $H = -\infty$). This can of course be studied numerically, and we shall display later on some results obtained in this way; in the next subsection we shall argue that, with some additional hypotheses on the nature of the transition one can compute its location analytically.

\subsection{Locating the transition}\label{sec:BRW}

Let us assume that the quantities $p$, $\hat p$ and $c$ defining the model depend on some continuous parameter denoted $\lambda$, and that there exists a threshold value $\bar\lambda$ such that $\mathbb{E}[\varrho](\lambda)=0$ for $\lambda>\bar\lambda$, and $\mathbb{E}[\varrho](\lambda)>0$ for $\lambda<\bar\lambda$. We further assume that the transition at $\bar\lambda$ is continuous, i.e. $\mathbb{E}[\varrho](\lambda)\to 0$ as $\lambda \to \bar\lambda^-$. Under these hypotheses, the expression \eqref{errore} of $\mathbb{E}[\varrho]$ reveals that $\mathbb{P}[\hat H+\hat H'>\hat\Omega]\to 1$ when $\lambda$ reaches its threshold value from the partial recovery phase. But if $\mathbb{P}[\hat H+\hat H'>\hat\Omega]=1$ the minimum in \eqref{cavmapmatch2} is always realized by the first argument, which leads us to study the following, simplified form of the RDE \eqref{cavmapmatch}:
\begin{subequations}\label{HH}
\begin{align}
\hat K&\stackrel{\mathrm{d}}{=}\min_{1\leq r\leq Z}\left[\Omega_{r}-K_{r}\right] \ , \\
  K&\stackrel{\mathrm{d}}{=}\hat \Omega-\hat K \ ,
\end{align}
\end{subequations}
which bear on a new couple of random variables $\hat K$ and $K$. The transition point will be characterized by the fact that the simplified RDE in Eqs.~\eqref{HH} has a non-trivial solution at $\bar \lambda$. To facilitate the discussion we define a new random variable
\begin{equation} \label{eq_def_Xi}
\Xi \stackrel{\mathrm{d}}{\coloneqq} \Omega-\hat\Omega \ ,
\end{equation}
in terms of which Eqs.~\eqref{HH} can be written as a distributional equation for $\hat K$ only,
\begin{equation}\label{eqsimp}
\hat K\stackrel{\mathrm{d}}{=}\min_{1\leq r\leq Z}\left[\Xi_{r}+\hat K_r\right] \ .
\end{equation}

This RDE is actually connected to the properties of Branching Random Walk (BRW) processes, a subject that has generated a vast literature both in physics and mathematics~\cite{Hammersley1974,Kingman1975,Biggins1977,Brunet1997,Majumdar2000,Ebert2000,Bachmann2000,Aldous2005,Aidekon2013,Shi15,Bramson2016}, in the more general context of front propagation for reaction-diffusion equations. For the convenience of the reader we summarize here the definitions and the main properties we need about BRWs, some more details can be found in Appendix~\ref{app:rde}. A BRW describes the evolution of a population of particles that move along a continuous unidimensional spatial axis, and multiply as time increases in discrete steps (the equivalent process in continuous time being the Branching Brownian Motion). More explicitly, at the initial generation $n=0$ there is a single particle at the origin, $X^{(0)}_1=0$. Each generation $n$ is given by a set of particles in positions $\{X_k^{(n)}\}_k$ and is constructed iteratively. Each particle of the generation $n$, say the $i$-th one at position $X_i^{(n)}$, gives rise to a number (possibly infinite) of offsprings in the next generation, located at the positions $X^{(n+1)}_{i,r}=X_i^{(n)}+\Xi_{i,r}$ where the displacements $\{\Xi_{i,r} \}_r$ between the positions of a parent particle and its offsprings are, independently for each $i$, copies of an identical point process. In the simplest cases the number of offsprings is $Z_i$, an independent copy of the random variable $Z$ for each parent $i$, and the displacements $\Xi_{i,r}$ are i.i.d. copies of a given random variable $\Xi$. 

A realization of such a process is pictured on the example below:
\begin{center}
\begin{tikzpicture}[scale=0.3,rotate=-90]
\draw[gray,-latex] (0,-5) -- node[pos=0,rotate=0,fill=white] {\footnotesize $n=0$} (0,8);
\draw[gray,-latex] (4,-5) -- node[pos=0,rotate=0,fill=white] {\footnotesize $n=1$} (4,8);
\draw[gray,-latex] (8,-5) -- node[pos=0,rotate=0,fill=white] {\footnotesize $n=2$} (8,8);
\draw[gray,-latex] (12,-5) -- node[pos=0,rotate=0,fill=white] {\footnotesize $n=3$} (12,8);
\node[circle,draw,inner sep=1.5pt,fill=black,label=above:{\footnotesize $X_1^{(0)}$}](o) at (0,0){};
\node[circle,draw,inner sep=1.5pt,fill=white](1) at (4,3){}; \draw (1) -- (o);
\node[circle,draw,inner sep=1.5pt,fill=white](2) at (4,-2){}; \draw (2) -- (o);
\node[circle,draw,inner sep=1.5pt,fill=white](11) at (8,7){}; \draw (11) -- (1);
\node[circle,draw,inner sep=1.5pt,fill=white](12) at (8,1){}; \draw (12) -- (1);
\node[circle,draw,inner sep=1.5pt,fill=white](13) at (8,0){}; \draw (13) -- (1);
\node[circle,draw,inner sep=1.5pt,fill=white](21) at (8,2){}; \draw (21) -- (2);
\node[circle,draw,inner sep=1.5pt,fill=white](22) at (8,-1){}; \draw (22) -- (2);
\node[circle,draw,inner sep=1.5pt,fill=white](111) at (12,5){}; \draw (111) -- (11);
\node[circle,draw,inner sep=1.5pt,fill=white](112) at (12,2){}; \draw (112) -- (11);
\node[circle,draw,inner sep=1.5pt,fill=white](121) at (12,3){}; \draw (121) -- (12);
\node[circle,draw,inner sep=1.5pt,fill=white](122) at (12,-3){}; \draw (122) -- (12);
\node[circle,draw,inner sep=1.5pt,fill=white](123) at (12,0){}; \draw (123) -- (12);
\node[circle,draw,inner sep=1.5pt,fill=white](131) at (12,1){}; \draw (131) -- (13);
\node[circle,draw,inner sep=1.5pt,fill=white](211) at (12,4){}; \draw (211) -- (21);
\node[circle,draw,inner sep=1.5pt,fill=white](212) at (12,6){}; \draw (212) -- (21);
\node[circle,draw,inner sep=1.5pt,fill=white](213) at (12,-1){}; \draw (213) -- (21);
\node[circle,draw,inner sep=1.5pt,fill=white](222) at (12,-2){}; \draw (222) -- (22);
\end{tikzpicture}
\end{center}
Note that BRWs combine and generalize Galton-Watson branching processes, that are recovered if one only looks at the number of particles in the BRW and discards their positions, and unidimensional random walks: the positions of the particles along a single branch of the BRW follow the law of a simple random walk

Among several properties of BRWs one that has attracted a lot of research effort is the asymptotic behavior in the large $n$ limit of the minimum of the process. Let us denote $\hat K^{(n)}=\min_{k}[X_k^{(n)}]$ the position of the leftmost particle in the $n$-th generation. Decomposing a BRW of depth $n+1$ into $Z$ BRW of depth $n$ attached to the root via $Z$ displacements $\Xi_{r}$ it is easy to convince oneself that $\hat K^{(n)}$ obey the following RDE,
\begin{equation}\label{simpitera}
\hat K^{(n+1)} \stackrel{\mathrm{d}}{=}\min_{1\leq r\leq Z}\left[\Xi_{r}+\hat K^{(n)}_r\right] \ ,
\end{equation}
with the initial condition $\hat K^{(0)}=0$ and the convention $\hat K^{(n)} = -\infty$ if the process is extinct before the $n$-th generation. Such sequences of random variables have been extensively studied, and very precise mathematical results have been obtained. A first level of description \cite{Hammersley1974,Kingman1975,Biggins1977} shows that, conditional on the non-extinction of the process, $\hat K^{(n)}$ has a ballistic behavior, namely $\hat K^{(n)}/n \xrightarrow[n\to+\infty]{\mathrm{a.s.}} v$, with an almost sure convergence towards a velocity $v$ that can be computed in terms of the law of the point process of the displacements:
\begin{equation}
v=-\inf_{\theta>0}\frac{1}{\theta}\ln\mathbb E\left[\sum_{r}\e^{-\theta\Xi_r}\right] \ .
\end{equation}
When the number of offspring is a random variable of law $Z$ and the displacements i.i.d. copies of $\Xi$ this simplifies into:
\begin{equation}\label{vgen}
v=-\inf_{\theta>0}\frac{1}{\theta}\ln\left( \mathbb E [Z]  \mathbb E\left[\e^{-\theta\Xi}\right] \right)\ ,
\end{equation}
see also Appendix~\ref{app:rde} for an heuristic justification of this expression of the velocity. More recently a finer description of the limit has been obtained~\cite{Bachmann2000,Aidekon2013,Bramson2016}: under some technical conditions there exists a constant $C$, such that, conditional on the non-extinction of the process,
\begin{equation}\label{asintoK}
\hat K^{(n)}- n v -C \log n\xrightarrow[n\to+\infty]{\mathrm{d}}L,
\end{equation}
where the convergence is in law and $L$ is a finite random variable satisfying
\begin{equation} \label{eq_BRW_Y}
L \stackrel{\mathrm{d}}{=} - v + \min_{1\leq r\leq Z}\left[\Xi_{r}+ L_r\right] \ .
\end{equation}
Note that this equation is invariant by translation: if $L$ is a solution then $L+a$ also is, for any constant $a$. Moreover the left tail behavior of the limit random variable $L$ was established in~\cite{Aidekon2013,Bramson2016} to be
\begin{equation}
\mathbb{P}[L \le z]\sim \alpha \, z\e^{\theta_* z} \quad \text{as} \ \ z\to - \infty \ ,
\label{eq_left_tail_Y}
\end{equation}
where $\theta_*$ is the minimizer of \eqref{vgen}, and $\alpha<0$ a constant.

Let us now come back to the planted matching problem, and specialize these results taking for the law $Z$ of the offspring of the BRW the expression \eqref{disZhat}, and for $\Xi$ the random variable defined in \eqref{eq_def_Xi}, where we recall that $\Omega$ (resp. $\hat \Omega$) is the random variable $\omega(W)$ with the function $\omega$ of \eqref{omega} and $W$ drawn with the distribution $P$ (resp. $\hat P$). Note that in our case $Z\ge 1$ with probability 1, hence the probability of extinction of the BRW process is equal to zero. The expression of the velocity \eqref{vgen} becomes then
\begin{equation}\label{ourvelocity}
v =-\inf_{\theta>0}\frac{\ln\left[I(\theta)I(1-\theta)\right]}{\theta} \ ,
\end{equation}
with
\begin{equation}
I(\theta)=\sqrt\gamma\int_\Gamma \hat P(w)^{\theta} P(w)^{1-\theta}\dd w \ .
\end{equation}
where we remind the definition of $\Gamma$ in Eq.~\eqref{def:gamma}.
One realizes at this point that if the parameters of the problem are such that $v=0$, then the random variable $L$ solution of \eqref{eq_BRW_Y} and constructed through the large generation limit of the BRW is a non-trivial solution of Eq.~\eqref{eqsimp}: this is precisely the condition we argued to be satisfied at the continuous phase transition between full and partial recovery phases. The vanishing velocity criterion
\begin{equation}\label{condizionegen}
\inf_{\theta>0}\frac{\ln\left[I(\theta)I(1-\theta)\right]}{\theta}=0
\end{equation}
is actually equivalent to $I(1/2)=1$ because the argument of the logarithm is a convex function symmetric around $\theta=1/2$, and therefore this condition becomes 
\begin{equation}\label{condizione}
\int_\Gamma\sqrt{\hat P(w)P(w)}\dd w=\frac{1}{\sqrt\gamma} \ ,
\end{equation}
or equivalently in terms of the original parameters:
\begin{equation}\label{condizione_original}
\int_\Gamma\sqrt{\hat p(w) p(w)}\dd w=\frac{1}{\sqrt c} \ .
\end{equation}
Note that the Cauchy-Schwarz inequality implies 
\begin{equation}
\int_{\Gamma}\sqrt{\hat P(w)P(w)}\dd w\leq 1 \ ,
\end{equation}
hence \eqref{condizione} cannot be satisfied if $\gamma < 1$. This is perfectly consistent with what we found in Sec.~\ref{sec_second_pruning}, if $\gamma < 1$ the only solution to \eqref{eq_q} is $q=1$, signalling a phase where full recovery can be achieved by the leaf removal procedure.

The equation \eqref{condizione_original} is our first main result; it provides a prediction for the locus of the continuous phase transition in the parameter space $(p,\hat p,c)$ of the model. We shall simplify it in the large degree limit in Sec.~\ref{sec_criterion_cinfty}, and test it numerically on several examples in Sec.~\ref{sec:examples}. Before that we shall make a series of remarks on the reasoning which led to it and on its consequences.

(i) We have implicitly assumed that $v=0$ is a necessary condition for \eqref{eqsimp} to have a non-trivial solution, in other words that the solution of \eqref{eq_BRW_Y} is unique (modulo the invariance under translations) and can thus be realized as the (properly shifted) large $n$ limit of the BRW construction. This uniqueness is actually an open question in mathematics, stated as open problem 46 in~\cite{Aldous2005}. 

(ii) We justified the introduction of the simplified RDE \eqref{HH} by an assumption on the continuity of $\mathbb{E}[\varrho]$. We can be more precise in some cases; suppose that the random variable $\hat\Omega$ is not bounded from above, i.e. that $\omega(w)$ diverges to $+\infty$ at some point in $\Gamma$ (as we will see later on there are non-trivial examples where this property can be true, or false). Then a continuously vanishing $\mathbb{E}[\varrho]$ in Eq.~\eqref{errore} can only occur if $\hat H$ diverges to $+\infty$ as $\lambda \to \bar\lambda^-$. In that case we can restate more precisely our hypotesis as the existence of a function $m(\lambda)$ that diverges to $+\infty$ as $\lambda \to \bar\lambda^-$, such that
\begin{equation}
\hat H - m(\lambda) \xrightarrow[\lambda \to \bar\lambda^-]{\mathrm{d}}\hat{K} \ ,
\label{eq_convergence_hatH}
\end{equation}
with $\hat{K}$ solution of \eqref{eqsimp}. The case studied in \cite{Moharrami2019} falls in this category, and Sec.~\ref{sec:critical} will be devoted to the determination of the divergence of $m(\lambda)$.

(iii) Independently of the continuity assumption, a point in parameter space with $v>0$ is most certainly in a full recovery phase, according to the following reasoning. Instead of the fixed point condition \eqref{cavmapmatch} consider an iterative version of these equations,
\begin{subequations} \label{eq_RDE_iterations}
\thinmuskip=0mu
 \begin{align}
  \hat H^{(n+1)}&\stackrel{\mathrm{d}}{=}\min_{1\leq i\leq Z}\left[\Omega_{i}-H_{i}^{(n)}\right]\ ,  \label{eq_RDE_iterations_a} \\
  H^{(n)}&\stackrel{\mathrm{d}}{=}
  \begin{cases}
  \hat \Omega-\hat H_1^{(n)}&\text{ with prob. $q$}\ , \\
  \min\left(\hat \Omega-\hat H_1^{(n)}, \ \hat H_2^{(n)}  \right)&\text{ with prob. $1-q$} \ ,
  \end{cases}  \label{eq_RDE_iterations_b}
 \end{align}
\end{subequations}
that defines a sequence of random variables $\hat H^{(n)}$, with the initial condition $\hat H^{(0)}=0$. Comparing these equations with \eqref{simpitera} one can show by induction on $n$ that $\hat K^{(n)}$ is stochastically smaller~\cite{Lindvall2012} than $\hat H^{(n)}$; we detail this proof in Appendix~\ref{appendix_ordering}.  Here we shall only recall that given two random variables $X$ and $Y$ one says that $X$ is stochastically smaller than $Y$, to be denoted $X \preceq Y$, if and only if $\mathbb{P}[X>x] \le \mathbb{P}[Y>x] $ for all $x$. This condition is equivalent to the existence of a coupling $(\hat X,\hat Y)$, i.e. a random vector with marginal laws equal to those of $X$ and $Y$ respectively, such that $\mathbb{P}[\hat{X} \le \hat Y]=1$. In our case if $v>0$ we have seen that $\hat K^{(n)}$ diverges to $+\infty$ in the large $n$ limit, hence by this stochastic comparison argument this will also be the case of $\hat H^{(n)}$. With the assumption that a non-trivial solution of the fixed point equation \eqref{cavmapmatch}, if it exists, will be reached as the large $n$ limit of the sequence $\hat H^{(n)}$, this allows to conclude that $v>0$ rules out such a non-trivial fixed point, hence is a criterion for a full recovery phase.

\subsection{The large degree limit}
\label{sec_criterion_cinfty}

As explained in Sec.~\ref{sec:definitions} the large degree limit, $c \to \infty$ taken here after the thermodynamic limit $N \to \infty$, allows to recover the dense models defined on complete graphs in~\cite{Chertkov2010,Moharrami2019}, if one performs an appropriate rescaling of the distribution of the weights on the non-planted edges. Consider indeed the condition \eqref{condizione_original}: if $p$ and $\hat p$ are kept constant as $c\to \infty$ this becomes $\int_\Gamma \sqrt{\hat p(w)p(w)}\dd w=0$, which is never satisfied unless $\Gamma$ is empty (and in this case the planted edges can be identified by inspection of the weight on the edges, as $p$ and $\hat p$ have disjoint supports). Indeed if the weights on both types of edges are of the same order of magnitude, around one vertex the planted weight will be hidden among the $O(c)$ non-planted ones, and impossible to distinguish in the $c \to \infty$ limit. To have a nontrivial partial-full recovery transition for $c\to \infty$ it is therefore necessary to scale the non-planted edges weights, the simplest way being to take the non-planted weight distribution uniform on the interval $[0,c]$. The condition \eqref{condizione_original}  becomes then
\begin{equation}
 \int\sqrt{\hat p(w)}\, \dd w=1 \ ,
\end{equation}
where we assumed for simplicity of notation that the support of $\hat{p}$ is included in the positive real axis.

\section{Examples}
\label{sec:examples}
We shall now confront our analytical prediction for the location of the phase transition with numerical results, obtained both for finite $N$ by solving the BP equations on single samples of the problem, and in the thermodynamic limit by solving numerically the RDEs. Some details on these numerical procedures are given in Sec.~\ref{sec:resolution}.

For concreteness we will always take an uniform distribution for the non-planted weights,
\begin{equation}\label{uniform}
p(w)=\frac{1}{c}\mathbb I(0 \le w \le c) \ ,
\end{equation}
for which Eq.~\eqref{condizione_original} further simplifies as
\begin{equation}
\int_\Gamma\sqrt{\hat p(w)}\, \dd w=1 \ ,\quad \Gamma=\supp(\hat p)\cap[0,c] \ .
\label{eq_criterion_uniform}
\end{equation}
We will present our results for different choices of $\hat p$, some of them partially investigated in the literature.

\subsection{The exponential distribution}\label{sec:exponential}
Let us start by considering the exponential distribution
\begin{equation}\label{expdis}
    \hat p(w)=\lambda \e^{-\lambda w}\mathbb{I}(w \ge 0) \ ,
\end{equation}
for which, in the large degree limit, \cite{Moharrami2019} proved that $\lambda < 4$ is a partial recovery phase, while $\lambda > 4$ corresponds to full recovery.

\begin{figure}
  \includegraphics[width=\columnwidth]{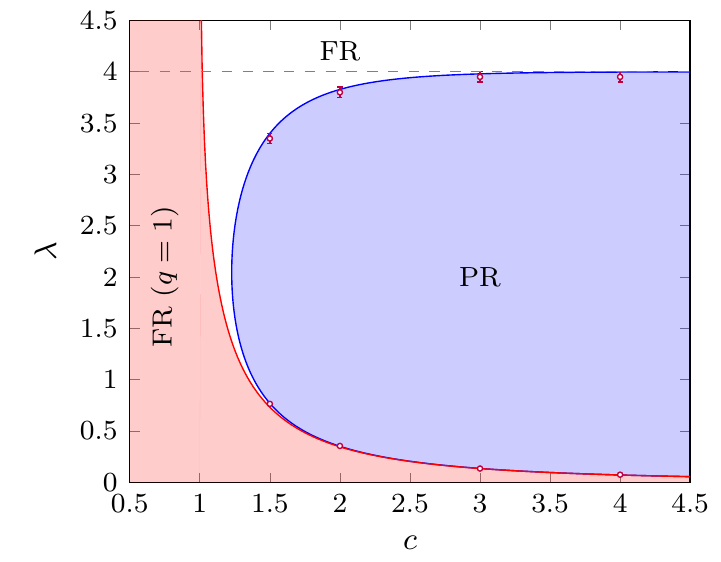}
\caption{Phase diagram of the planted matching problem with exponential planted weights on sparse graphs, exhibiting full-recovery (FR) and partial recovery (PR) phases. The red area corresponds to the phase where full recovery is achievable by the simple leaf removal procedure, and is delimited by the $\gamma=1$ condition (red line). The blue area is the partial recovery phase enclosed by the vanishing velocity criterion (blue line), the white domain corresponding to full recovery. The red dots have been obtained from the numerical resolution of the RDEs \eqref{cavmapmatch} by a population dynamics algorithm with $10^6$ fields, and mark the limit of existence of a non-trivial solution.
\label{fig:phased}}
\end{figure}

Our predictions are summarized in the phase diagram in the $(c,\lambda)$ plane displayed in Fig.~\ref{fig:phased}. The red line corresponds to the condition $\gamma=1$ below which the leaf removal procedure described in Sec.~\ref{sec_second_pruning} recovers completely the hidden matching; here $\mu=1$ and $\hat \mu = 1-\e^{-\lambda c}$, the equation of this line is thus $\lambda = - \log(1-(1/c))/c$. The blue line is instead the vanishing velocity criterion \eqref{eq_criterion_uniform}, which becomes for this choice of $\hat p$:
\begin{equation}
1=\int_0^c\sqrt\lambda\e^{-\frac{\lambda w}{2}}\dd w=2\frac{1-\e^{-\frac{c\lambda }{2}}}{\sqrt\lambda} \ ,
\end{equation}
a relation that can be inverted in
\begin{equation}
 c=-\frac{2}{\lambda}\ln\left(1-\frac{\sqrt{\lambda}}{2}\right) \ .
\end{equation}
This blue line separates a domain, in blue in Fig.~\ref{fig:phased}, where $v<0$, corresponding to a partial recovery phase, from a full recovery phase (in white) with $v>0$. Note that the curve has a minimal abscissa of $c \approx 1.2277$ below which there is full recovery for all $\lambda$. In the large degree limit the two branches of the blue line converge to $\lambda =0$ and $\lambda =4$, we thus recover the results of~\cite{Moharrami2019}.

\begin{figure}
 \includegraphics[width=\columnwidth]{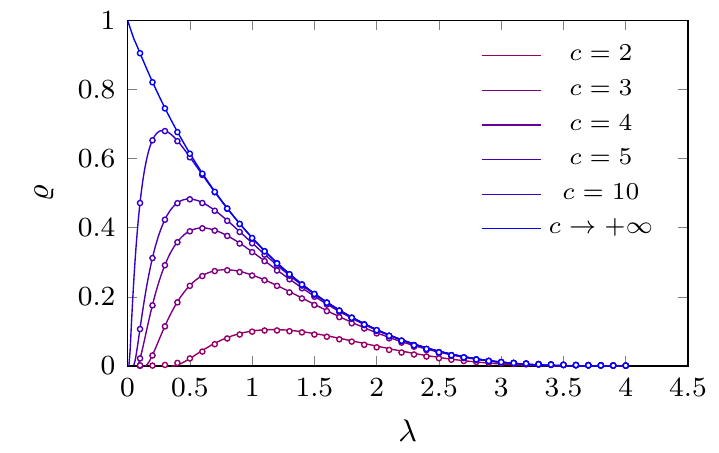}
\caption{Reconstruction error for the planted matching problem with exponential planted weights and different values of $c$. The lines have been obtained numerically solving the RDEs \eqref{cavmapmatch} using a population dynamics algorithm with $10^6$ fields. The dots are the results of the resolution of the BP equations \eqref{cav1} on graphs of $N=10^3$ vertices, averaged over $10^4$ instances, the $c\to+\infty$ case corresponding to the complete graph.
\label{fig:exp}}
\end{figure}

The red dots on this phase diagram have been obtained from a numerical resolution of Eqs.~\eqref{cavmapmatch}, using a population dynamics algorithm, and correspond to the limit values for which we found a non-trivial solution of the equations. They are in agreement, within numerical accuracy, with the analytical prediction. In Fig.~\ref{fig:exp} we compare our prediction for the average reconstruction error $\mathbb{E}[\varrho]$ (non-zero in the partial recovery phase) obtained by Eqs.~(\ref{cavmapmatch}, \ref{errore}) (in the thermodynamic limit) with the numerical results obtained running a belief propagation based on Eq.~\eqref{cav1} (on finite graphs). The agreement between these two procedures is very good, except for smoothening finite-size effects close to the phase transitions.

\subsection{The folded Gaussian case}\label{sec:gauss}
\begin{figure}
 \includegraphics[width=\columnwidth]{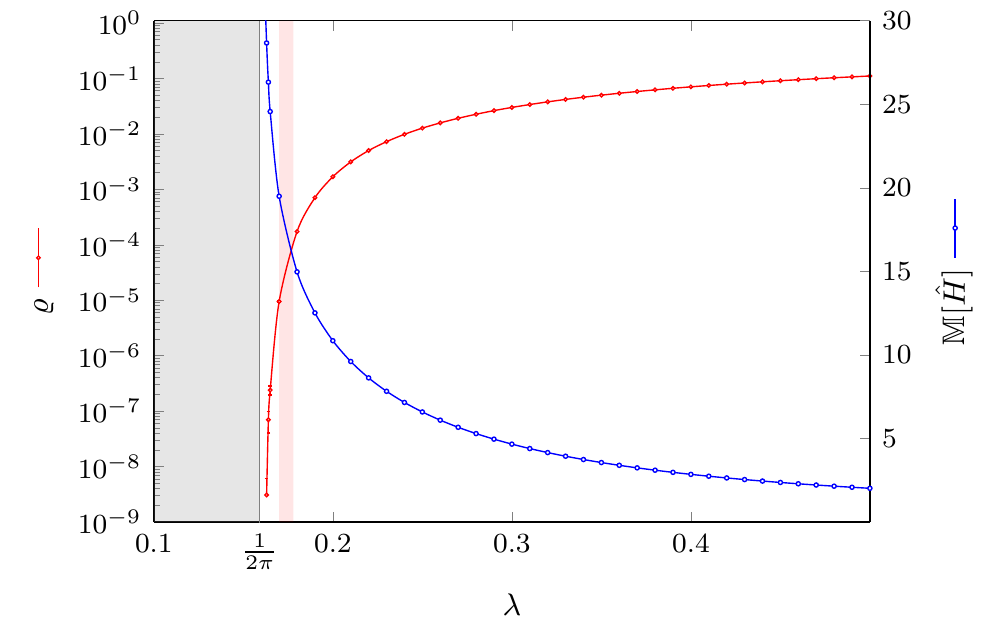}
\caption{Planted matching problem with (folded) Gaussian weights, in the large degree limit; the curves have been obtained from a numerical resolution of Eqs.~\eqref{AAhat}, the red line corresponding to the reconstruction error $\varrho$, the blue line to the median cavity fields on planted edges $\mathbb M[\hat H]$. The light-red interval corresponds to the estimate for the transition point given in Ref.~\cite{Chertkov2010}.
\label{fig:gauss}}
\end{figure}

We have also considered a planted weight distribution of the folded Gaussian form,
\begin{equation}
\hat p(w)=\sqrt{\frac{2}{\pi \lambda}}\e^{-\frac{ w^2}{2\lambda}}\mathbb{I}(w \ge 0) \ ,
\end{equation}
that had been investigated previously in~\cite{Chertkov2010} in the large degree limit. Our prediction for this case, easily obtained by plugging this expression in \eqref{eq_criterion_uniform} and taking the $c \to \infty$ limit, is of a phase transition at 
\begin{equation}
 \bar\lambda=\frac{1}{2\pi} \approx 0.1591
\end{equation} 
between a partial recovery phase for $\lambda>\bar\lambda$ and a full-recovery phase for $0<\lambda<\bar\lambda$. This agrees qualitatively with the numerical investigations of~\cite{Chertkov2010}, but not with the value of the threshold that was estimated in~\cite{Chertkov2010} to be $\bar\lambda=0.174(4)$. We believe this discrepancy is due to finite-population size effects (that are particularly severe in this kind of problems, as discussed in Section~\ref{sec:resolution}) in the study of~\cite{Chertkov2010}. To check this point we solved the equations with a careful numerical integration (see Sec.~\ref{sec:resolution}) of the recursive equations, which convincingly suggests that $\bar\lambda<0.160(1)$, as shown on Fig.~\ref{fig:gauss}. 

\subsection{The truncated power-law case}\label{sec:powerlaw}
Let us consider as a final example the case where the density of the weights on the planted edges varies as a power-law on a finite interval,
\begin{equation}
 \hat p(w)=\frac{\alpha w^{\alpha-1}}{\lambda^{\alpha}}\mathbb I(0\leq w\leq \lambda),
\end{equation}
with $\alpha>0$. For the sake of simplicity, we will restrict our analysis to the case $c>\lambda$. Then one finds immediately that $\Gamma=[0,\lambda]$, $\hat P = \hat p$,  $P$ is the uniform distribution on  $[0,\lambda]$ and $\gamma=\lambda$, in such a way that as soon as $c>\lambda$ the problem on the pruned graph is completely independent of $c$.

\begin{figure}
  \includegraphics[width=\columnwidth]{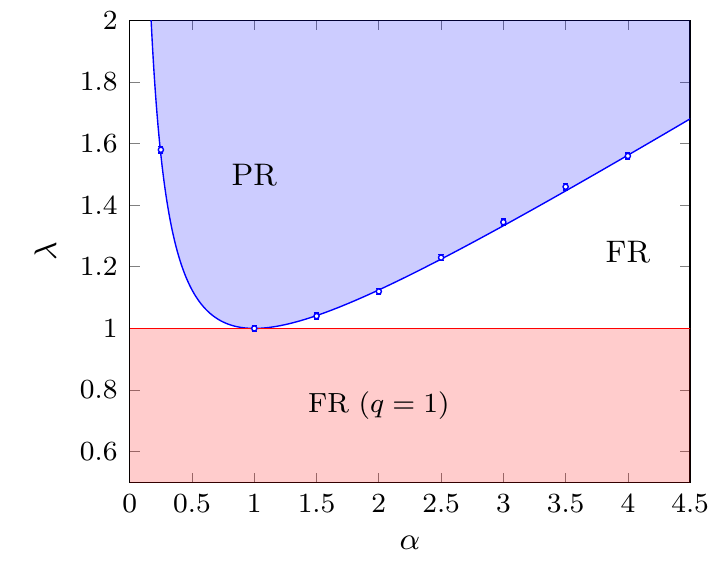}
\caption{Phase diagram of the recovery transition for the planted matching problem with a truncated power-law distribution for the planted weights. The red line corresponds to the $\gamma=1$ bound for the full-recovery transition. The circles are the transition points obtained running a population dynamics algorithm with $10^6$ fields. \label{fig:phasedpow}}
\end{figure}

The transition condition $\gamma=1$ for the full recoverability of the planted matching by the leaf removal algorithm is thus $\lambda=1$, which yields the red domain in the phase diagram of Fig.~\ref{fig:phasedpow}. The vanishing velocity condition \eqref{eq_criterion_uniform} becomes here
\begin{equation}\label{lampow}
\bar\lambda=\frac{(1+\alpha)^2}{4\alpha} \ ,
\end{equation}
which is plotted as a blue line in Fig.~\ref{fig:phasedpow}, the full-recovery phase corresponding to the domain $\lambda<\bar\lambda$. This prediction is in agreement with the numerical results obtained by a population dynamics resolution of the RDEs.

For $\alpha=1$, that corresponds to the planted weights uniformly distributed on $[0,\lambda]$, one can actually solve the RDEs explicitly. Indeed in this case the function $\omega(w)$ in Eq.~\eqref{omega} vanishes, hence the equations \eqref{cavmapmatch} reduce to
\begin{subequations}
\thinmuskip=0mu
 \begin{align}
  \hat H&\stackrel{\mathrm{d}}{=}-\max_{1\leq i\leq Z}\left[H_{i}\right]\ , \\
  H&\stackrel{\mathrm{d}}{=}
  \begin{cases}
  -\hat H&\text{ with prob. $q$} \ ,\\
  \min\left(-\hat H , \ \hat H'\right)&\text{ with prob. $1-q$}\ ,
  \end{cases} 
 \end{align}
\end{subequations}
which obviously admit the solution $H \stackrel{\mathrm{d}}{=}\hat H \stackrel{\mathrm{d}}{=} 0$. Indeed the effective weights on the pruned graph are all equal, the planted matching is one of the many perfect matchings of this reduced graph, but there is no information contained in the weights to decide which one. In a simple-minded application of the inclusion rule \eqref{eq_inclusion_rule} one would include in the estimator the edges of the pruned graph independently with probability $1/2$, leading to an average estimation error $\mathbb{E}[\varrho]=(1-q)^2(1+\lambda)/4$ for $\lambda > 1$, and of course $\mathbb{E}[\varrho]=0$ for $\lambda \le 1$.

\subsection{A note on the numerical procedures}\label{sec:resolution}

Most of the thermodynamic limit results presented above have been obtained by a numerical resolution of Eqs.~\eqref{cavmapmatch} via a population dynamics algorithm~\cite{Mezard2001}. The idea of this method, which is very commonly used to solve RDEs, is to represent the law of a random variable $X$ as the empirical distribution of a sample $\{X_1,\dots,X_{\mathzapf N}\}$ of its representants, with $\mathzapf N \gg 1$ to improve the accuracy of the method. In terms of cumulative distributions this corresponds to the approximation
\begin{equation}
F_X(x) \approx \frac{1}{\mathzapf N} \sum_{i=1}^{\mathzapf N} \mathbb{I}(X_i \le x) \ .
\label{eq_population}
\end{equation}
One considers then the iterative version of the RDE written in \eqref{eq_RDE_iterations}, and update the population according to these rules. For instance each representant $\hat H_i$ at the iteration $n+1$ is generated, independently, by drawing an integer $Z$ with the law \eqref{disZhat}, $Z$ copies of the random variable $\Omega$, and $Z$ representants of $H$ at the iteration $n$, by a uniform choice over the $\mathzapf N$ ones. These quantities are then combined according to the right hand side of \eqref{eq_RDE_iterations_a} to compute $\hat H_i$. The sample of representants of $H$ at the iteration $n+1$ is then generated similarly according to \eqref{eq_RDE_iterations_b}. These steps are repeated a large number $n$ of times, the type of phase (partial or full recovery) is then decided according to the convergence or divergence to $+\infty$ of the population representing $\hat H^{(n)}$ in the large $n$ limit. The accurate determination of such a phase transition suffers from finite population size effects that are much more severe than in usual applications of the population dynamics algorithm. Indeed the transition is governed by an instability that manifests itself as a front propagation in the cumulative distribution function; such front propagations are generically driven by the behavior in the exponentially small tail far away from the front~\cite{Brunet1997,Majumdar2000,Ebert2000}. As the finite population size implies a cutoff of $1/\mathzapf N$ on the smallest representable value of the cumulative distribution function, this translates into logarithmic finite population size effects on the velocity of the front and the location of the phase transition, at variance with the usual $1/\mathzapf N$ corrections for the computation of observables as empirical averages. We refer the reader to~\cite{Brunet1997} for a quantitative study of these logarithmic corrections in the velocity of a front in presence of a threshold in its tail. 

We thus believe that the discrepancy in the folded Gaussian case between our analytical prediction $\bar\lambda=\frac{1}{2\pi} \approx 0.159$ and the numerical estimate $\bar\lambda=0.174(4)$ of~\cite{Chertkov2010} can be ascribed to these strong finite-$\mathzapf N$ effects. The results presented in Fig.~\ref{fig:gauss} that supports this thesis have been obtained with another numerical procedure: instead of the population representation \eqref{eq_population} of the cumulative distribution functions $F_H(h)$ and $F_{\hat H}(h)$ we stored their values in $M$ points $h_1<h_2<\dots<h_M$ over a given interval $[h_1,h_M]$, and updated them using Eqs.~\eqref{AAhat} until a certain convergence criterion was satisfied (until the $L^2$-distance between the solution at step $n$ and the solution at step $n-1$ was smaller than a given, pre-fixed tolerance $\epsilon$). The advantage of this method is that the cutoff $h_M$ can be taken arbitrarily large, in such a way that $\bar{F}_{\hat H}(h_M)$ is very small, hence bypassing the threshold at $1/\mathzapf N$ of the population dynamics algorithm. After convergence the function $F_{\hat H} (h)$ can be used to estimate $\mathbb{E}[\varrho]$, e.g. by a Monte Carlo integration. 

\section{A more precise description of the critical regime}
\label{sec:critical}

Once the threshold value of a parameter has been determined it is natural to aim at a more quantitative description of the transition in its critical regime. In the case considered in this paper of planted models that undergo a continuous transition from partial recovery for $\lambda < \bar \lambda$ to full recovery for $\lambda > \bar \lambda$ this point amounts to describe how the average reconstruction error $\mathbb{E}[\varrho]$ vanishes as $\lambda \to \bar \lambda^-$. This was raised as open question 2 in~\cite{Moharrami2019}, and we shall study it in the model defined therein, i.e. with an exponential distribution for the planted weights, in the large degree limit. This case allows for some technical simplifications; as shown in~\cite{Moharrami2019} the RDEs can then be reduced to a system of Ordinary Differential Equations (ODEs), that we first recall in the next subsection before studying their solution in the critical regime.

\subsection{The ODEs for the exponential model}

Let us specialize our formalism with the following choices of weight distributions: $\hat{p}(w) = \lambda\e^{-\lambda w}$ for $w \ge 0$, and $p(w)=1/c$ for $w \in [0,c]$. The intersection of their supports is thus $\Gamma=[0,c]$, and one finds $\mu=1$, $\hat \mu=1-\e{-\lambda c}$. The reduced distributions $\hat P$ and $P$ are then, on their common support $\Gamma$, $\hat{P}(w)=\lambda \e^{-\lambda w}/\hat \mu$ and $P(w)=1/c$, which gives an effective weight function $\omega(w) = \lambda w -\ln(\lambda c /\hat \mu)$. The parameter $q$ is the solution of $q=\e^{-\gamma (1-q)}$ with $\gamma = c \hat \mu$.

To simplify the notations, and to get closer to the conventions used in~\cite{Moharrami2019}, we shall define random variables $X$ and $Y$ that are affine transformations of $\hat H$ and $H$, respectively. More precisely we define their cumulative functions as
\begin{align}
F_X(x)&\coloneqq  F_{\hat H}\left(\lambda \, x-\frac{1}{2} \ln\left( \frac{ \lambda c}{\hat \mu} \right)\right) \ , \\
F_Y(x) &\coloneqq  F_H\left(\lambda \, x-\frac{1}{2} \ln\left( \frac{ \lambda c}{\hat \mu} \right)\right) \ ,
\end{align}
and keep the convention $\bar{F}=1-F$ for reciprocal cumulative distributions.
The equations \eqref{AAhat} become
\begin{subequations}\label{AAhexpC}
 \begin{align}
\bar{F}_Y(x)&=\left(q+(1-q)\bar{F}_X(x)\right)\frac{\lambda}{\hat\mu} \int_0^c\e^{-\lambda w} F_X(w-x)\dd w \\
\bar{F}_X(x)&=\frac{\exp\left[-\hat\mu(1-q)\int_{-x}^{c-x} \bar{F}_Y(w)\dd w\right]-q}{1-q}.
 \end{align} 
\end{subequations}
Taking the limit $c\to+\infty$, in which $\hat\mu\to 1$ and $q\to 0$, Eqs.~\eqref{AAhexpC} become
\begin{subequations}\label{AAhexp}
 \begin{align}
\medmuskip=0mu
\thinmuskip=0mu
\thickmuskip=0mu
\bar{F}_Y(x)&= 
\bar{F}_X(x)\int_{0}^{+\infty}\lambda \e^{-\lambda w}F_X(w-x) \dd w \ ,\label{AAhexp1}\\
\bar{F}_X(x)&=\exp\left(-\int_{-x}^{+\infty} \bar{F}_Y(w)\dd w \right)\ . \label{AAhexp2} 
 \end{align} 
\end{subequations}
These equations between cumulative distribution functions correspond to the following RDEs on $X$ and $Y$:
\begin{subequations}
\begin{align}
X &\stackrel{\mathrm d}{=} \min \, \{ \xi_i - Y_i \} \ ,\\
Y &\stackrel{\mathrm d}{=} \min (\eta-X,X') \ ,
\end{align}
\end{subequations}
where in the first line the $\xi_i$'s are the points of a Poisson point process of intensity 1 on the positive real axis, and in the second line $\eta$ has an exponential distribution of parameter $\lambda$. It is convenient to introduce the auxiliary function $V(x)$ defined as the cumulative distribution of the random variable $X-\eta$, i.e.
\begin{equation}
V(x) \coloneqq \mathbb{P}[X - \eta \le x] = \int_{0}^{+\infty}\lambda \e^{-\lambda w}F_X(w+x) \dd w \ ,
\label{eq_def_V}
\end{equation}
in such a way that \eqref{AAhexp1} can be rewritten $\bar{F}_Y(x)= \bar{F}_X(x) V(-x)$. Taking derivatives with respect to $x$ in (\ref{AAhexp2},\ref{eq_def_V}), and denoting for simplicity $F=F_X$, one obtains
\begin{subequations} \label{eq_FVderiv}
\begin{align}
F'(x)&=(1-F(x))(1-F(-x))V(x) \ , \label{eq_Fderiv} \\
V'(x) &= \lambda (V(x)-F(x)) \ ,
\end{align}
\end{subequations}
where the form of the equation on $V$ crucially depends on the exponential character of the distribution of the planted weight $\eta$. These two equations on $F$ and $V$ are not yet ODEs because one of the arguments in the right hand side of \eqref{eq_Fderiv} is $-x$ instead of $x$; to bypass this difficulty one introduces two additional functions, $G(x) \coloneqq F(-x)$ and $W(x) \coloneqq V(-x)$, in such a way that the four-dimensional vector $(F,G,V,W)(x)$ obeys an autonomous first-order ODE, from the solution of which the average reconstruction error is computed as
\begin{align}
\mathbb{E}[\varrho] &= \mathbb{P}[X+X' \le \eta] \nonumber \\
&= 2 \int_0^\infty (1-F(x))(1-G(x)) V(x) W(x) \dd x \ ,
\label{eq_rho_exponential}
\end{align}
see~\cite{Moharrami2019} for the details of the derivation of the integral expression of $\mathbb{E}[\varrho]$.

The dimensionality of the problem can be reduced by exploiting the conservation law $F(x)W(x)+G(x)V(x)-V(x)W(x)=0$ for all $x$. Introducing finally $U(x)=F(x)/V(x)$ it is shown in~\cite{Moharrami2019} that the problem reduces to solve, for $x \ge 0$, the following ODE on the three-dimensional vector $(U,V,W)(x)$:
\begin{subequations}\label{equvw}
\begin{align}
U'(x)  =&  - \lambda U(x) (1-U(x)) \label{eq_U} \\
& + (1-U(x)V(x))(1-(1-U(x))W(x)) \ , \nonumber \\
V'(x)  = & \lambda V(x) (1-U(x)) \ , \label{eq_V} \\
W'(x)  = & - \lambda W(x) U(x) \ , \label{eq_W} 
\end{align} 
\end{subequations}
with the initial conditions
\begin{equation}
U(0)=\frac{1}{2} \ , \qquad V(0)=W(0) \ .
\label{eq_boundary}
\end{equation} 
Even if the notation does not show it explicitly the solution of these ODEs depends of course on $\lambda$, both directly as $\lambda$ appears in \eqref{equvw}, and indirectly through the initial condition $V(0)=W(0)$. It is indeed shown in~\cite{Moharrami2019} that for a given $\lambda<4$ there is a unique choice of this initial condition that yields a proper solution, i.e. one in which $F$ and $V$ have the properties of cumulative distribution functions (non-decreasing and bounded between $0$ and $1$).

\subsection{The divergence of $X$ in the limit $\lambda \to 4$}

\begin{figure}
  \includegraphics[width=\columnwidth]{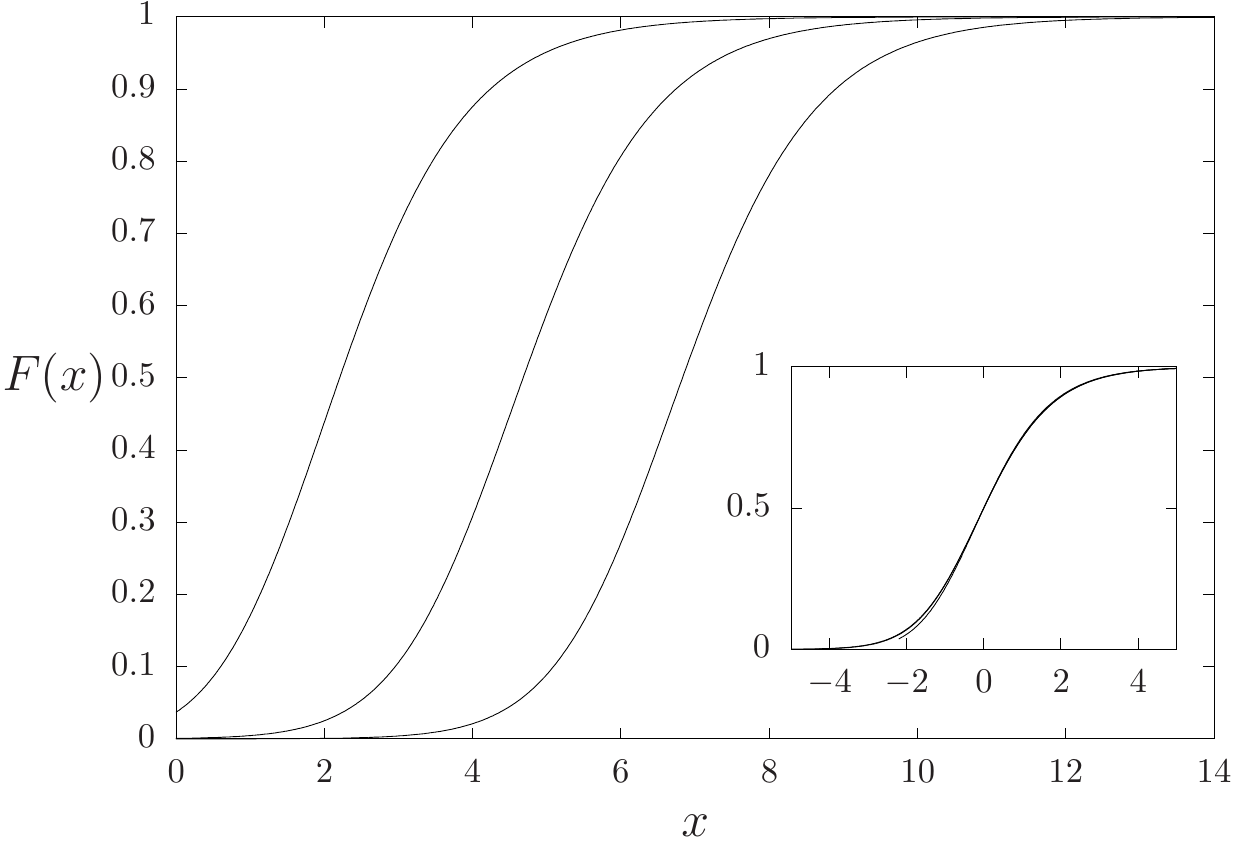}
\caption{Cumulative distribution $F(x)$ for, from left to right, $\lambda=3$, $\lambda=3.8$, $\lambda=3.92$. Inset: the three curves have been shifted horizontally by their medians $\mathbb{M}(\lambda) = F^{-1}(1/2)$, the collapse confirms the hypothesis stated in \eqref{eq_Xhat} of a convergence in distribution of the shifted random variables.
\label{fig:exp_F}}
\end{figure}

\begin{figure}
  \includegraphics[width=\columnwidth]{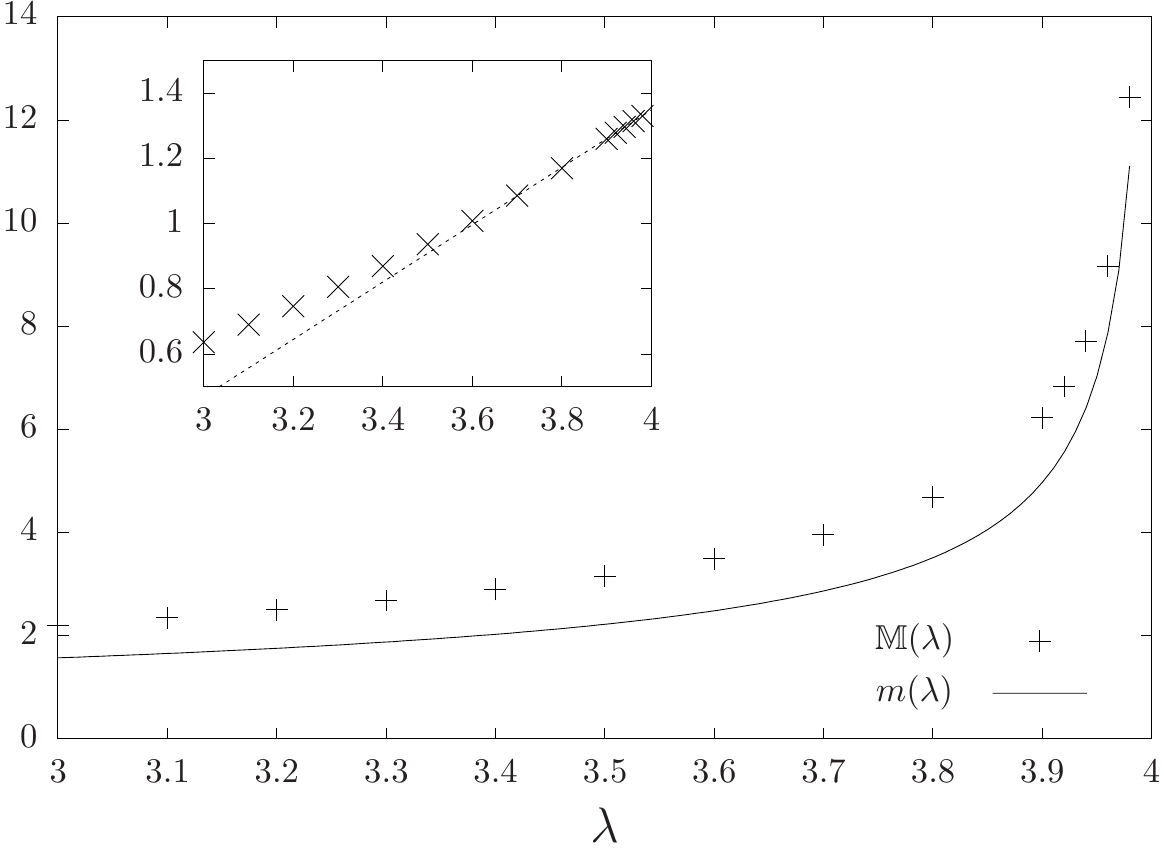}
\caption{Median $\mathbb{M}(\lambda)$ of the cumulative distribution in the exponential case, compared to the analytical prediction \eqref{eq_prediction_moflambda} for $m(\lambda)$. Inset: the dots show $\mathbb{M}(\lambda)-m(\lambda)$, the dashed line is a linear fit that confirms the convergence of $\mathbb{M}(\lambda)-m(\lambda)$ to a finite constant in the limit $\lambda \to 4$.
\label{fig:moflambda}}
\end{figure}

We present in Fig.~\ref{fig:exp_F} the cumulative distribution $F$ of the random variable $X$, for three values of $\lambda$ increasing towards the critical value $\lambda=4$, obtained by a numerical resolution of the ODE \eqref{equvw}. This plot suggests that $F$ drifts without deformation when approaching the transition; this impression is confirmed by the inset of the figure, which shows a very good collapse of the curves once shifted by the median $\mathbb{M}(\lambda)=F^{-1}(1/2)$ of $X$ (any other quantile would have led to the same collapse, with an additional constant shift of the horizontal axis). This observation has two equivalent translations: from the probabilistic point of view it corresponds to the existence of a function $m(\lambda)$ and a random variable $\hat X$ such that
\begin{equation}
X - m(\lambda) \xrightarrow[\lambda \to 4^-]{\mathrm{d}} \hat{X} \ ,
\label{eq_Xhat}
\end{equation}
as was stated in \eqref{eq_convergence_hatH} for generic weight distributions, $m(\lambda)$ differing from $\mathbb{M}(\lambda)$ by an arbitrary constant. From the analytic point of view it means that the solution of the ODEs admits a scaling regime $x=z+m(\lambda)$ when $z$ is kept fixed while $\lambda \to 4^-$, described by functions $\hat{F}(z)$, $\hat{V}(z)$, $\hat{U}(z)$, defined as 
\begin{equation}
\hat{F}(z) = \lim_{\lambda \to 4^-} F(z + m(\lambda)) \ , 
\end{equation}
similar definitions holding for $\hat{V}$ and $\hat{U}$. The dots in the main panel of Fig.~\ref{fig:moflambda} represent the numerically determined value of $\mathbb{M}(\lambda)$, and suggest a divergence of this quantity as $\lambda \to 4^-$. Supposing that $m(\lambda)$ indeed diverges one can simplify the ODEs \eqref{eq_FVderiv} in this scaling regime, with $F(-x) \to 0$; this yields
\begin{subequations} 
\begin{align}
\hat{F}'(x)&=(1-\hat{F}(x)) \hat{V}(x) \ ,  \\
\hat{V}'(x) &= 4 (\hat{V}(x)-\hat{F}(x)) \ .
\end{align}
\end{subequations}
$\hat{F}$ is the cumulative distribution of the limit variable $\hat{X}$, solution of the simplified RDE
\begin{subequations}
\begin{align}
\hat{X} &\stackrel{\mathrm d}{=} \min \, \{ \xi_i - \hat{Y}_i \} \ ,\\
\hat{Y} &\stackrel{\mathrm d}{=} \eta-\hat{X} \ .
\end{align}
\end{subequations}
Studying this simplified ODE in the $z \to - \infty$ limit where it can be linearized, or appealing to the theorems explained in \eqref{eq_left_tail_Y} for the left tail behavior of the limit random variable in the BRW interpretation, one finds
\begin{equation}
\hat{F}(z) \underset{z \to - \infty}{\sim} -A \, z \, \e^{2 z} \ , \quad
\hat{U}(z) \underset{z \to - \infty}{\sim}\frac{1}{2} - \frac{1}{4 z} \ , 
\label{eq_zminfty}
\end{equation}
where $A>0$ is a constant that cannot be fixed because of the invariance by translation of the equation.

In order to determine the sought-for divergence of $m(\lambda)$ as $\lambda \to 4^-$ we need now to study the solution of the ODE \eqref{equvw} on another scaling regime, $x = t m(\lambda)$ with $t \in [0,1)$ kept fixed in the limit, that will allow to take into account the initial condition \eqref{eq_boundary}. The derivation will then conclude by a matching argument at the common boundary of the two scaling regimes, $t \to 1^-$ and $z \to -\infty$. 

In order to study the scaling regime $x = t m(\lambda)$ we first notice that $V(0) \to 0$ as $\lambda \to 4^-$, because $V$ is the cumulative distribution function of the random variable $X - \eta$ that diverges in this limit. It is thus instructive to solve first Eqs.~\eqref{equvw} with $V(0)=W(0)=0$, denoting $U_0,V_0,W_0$  its solution, even if this cannot be exactly a proper solution. One finds $V_0(x)=W_0(x)=0$ for all $x \ge 0$, and \eqref{eq_U} simplifies into
\begin{equation}
U_0'(x)=-\lambda U_0(x)(1-U_0(x))+1 \ , 
\end{equation}
an equation that can be solved exactly for any $\lambda < 4$ with the initial condition $U_0(0)=1/2$, yielding
\begin{equation}
U_0(x)=\frac{1}{2}+\frac{1}{2}\sqrt{\frac{4-\lambda}{\lambda}} \tan \left(\frac{x}{2} \sqrt{\lambda(4-\lambda)} \right) \ .
\end{equation}
This expression diverges when the argument of the tangent reaches $\pi /2$, which gives us a natural candidate for the scale $m(\lambda)$ at which the first regime ends, namely
\begin{equation}
m(\lambda) = \frac{\pi}{2\sqrt{4 -\lambda}} \ ,
\label{eq_prediction_moflambda}
\end{equation}
and a conjecture for the behavior of the solution $U(x)$ of the full ODE in the first scaling regime, namely
\begin{equation}
\underset{\lambda \to 4^-}{\lim} \frac{1}{\sqrt{4-\lambda}} \left( U(t m(\lambda))-\frac{1}{2} \right) = 
\frac{1}{4} \tan\left(t \frac{\pi}{2} \right) \ .
\label{eq_scaling_U}
\end{equation}
We have assumed here that $V(0)$, even if strictly non-zero for all $\lambda <4$, is sufficiently small for the second line in \eqref{eq_U} to be negligible, and hence for $U$ to coincide with $U_0$ at the dominant order in this scaling regime. To check the self-consistency of this hypothesis we first give an exact expression of $V$ and $W$ in terms of $U$ obtained by integration of (\ref{eq_V},\ref{eq_W}):
\begin{align}
V(x) &= V(0) \exp\left[ \lambda \int_0^x (1-U(y)) \dd y  \right] \ , \label{eq_VofU} \\
W(x) & = \e^{-\lambda x} V(x) \ . \label{eq_WofV}
\end{align}
Inserting the scaling ansatz \eqref{eq_scaling_U} into \eqref{eq_VofU} we obtain 
\begin{equation}
\underset{\lambda \to 4^-}{\lim} \frac{1}{V(0)}V(t m(\lambda)) \e^{-2 t m(\lambda)} = \cos\left(t \frac{\pi}{2} \right) \ ,
\label{eq_scaling_V}
\end{equation}
the behavior of $W$ being easy to deduce from the one of $V$ thanks to \eqref{eq_WofV}. We fix now the initial condition $V(0)$ by matching the behavior $t \to 1^-$ of this expression with the limit $z \to -\infty$ of the other regime, which from \eqref{eq_zminfty} is $\hat{V}(z) \sim - 2 A z \e^{2 z}$, with the correspondance $t \sim 1 + \frac{z}{m(\lambda)}$. This yields
\begin{equation}
V(0) = 2 A \e^{-2 m(\lambda)} \frac{1}{\sqrt{4 - \lambda}} \ ,
\label{eq_scaling_V0}
\end{equation}
and allows to check that indeed the first line of \eqref{eq_U} is dominant in the scaling regime $t = m(\lambda)$ as long as $t<1$, confirming the self-consistency of our hypothesis. In addition the behavior of $U$ in \eqref{eq_zminfty} and \eqref{eq_scaling_U} matches at the boundary of the two scaling regimes. Note that the indeterminacy of the constant $A$, because of the invariance by translation of the equations on $\hat{F}$ and $\hat{V}$, is related in \eqref{eq_scaling_V0} to the additive arbitrary constant that can be added to $m(\lambda)$.

The inset of Fig.~\ref{fig:moflambda} presents a numerical confirmation of this reasoning: the difference between the numerically determined median of $X$ and our formula \eqref{eq_prediction_moflambda} is seen to converge to a finite constant when $\lambda \to 4^-$ (with corrections that seem polynomial in $4-\lambda$). We have also checked that the numerical results for $U(x)$ and $V(x)$ are compatible with the scaling ansatz of (\ref{eq_scaling_U},\ref{eq_scaling_V}).

\subsection{The critical behavior of $\mathbb{E}[\varrho]$}

We would like now to use our prediction \eqref{eq_prediction_moflambda} for the divergence of $X$ in order to describe the way in which the average reconstruction error $\mathbb{E}[\varrho]$ vanishes at the transition. The expression of the latter, given in \eqref{eq_rho_exponential}, can be rewritten as
\begin{equation}
\mathbb{E}[\varrho] = \mathbb{E}[\e^{-\lambda X}]^2 \ .
\end{equation}
Indeed the exponential distribution of $\eta$ is such that $\mathbb{P}[\eta \ge x]=\e^{-\lambda x}$, and $X$ and $X'$ in \eqref{eq_rho_exponential} are independent random variables with the same law. Recalling the convergence in distribution stated in \eqref{eq_Xhat}, and the prediction \eqref{eq_prediction_moflambda} of $m(\lambda)$, it would be tempting to write
\begin{equation}
\mathbb{E}[\varrho] \underset{\lambda \to 4^-}{\propto} \e^{- \frac{4 \pi}{\sqrt{4 -\lambda}} }\mathbb{E}[\e^{-4 \hat{X}}]^2 \ .
\end{equation}
Unfortunately this result has to be amended: as the tail of $\hat{X}$ varies as $\e^{2z}$ for $z \to -\infty$ (cf. Eq.~\eqref{eq_zminfty}) the expectation value $\mathbb{E}[\e^{-4 \hat{X}}]$ is infinite. Using the integral expression of $\mathbb{E}[\varrho] $ given in \eqref{eq_rho_exponential}, one finds that the leading contribution is given by the scaling regime $x=t m(\lambda)$ and is of the form
\begin{align}
2 &\int_0^{m(\lambda)} V(x) W(x) \dd x = 2 \int_0^{m(\lambda)} V(x)^2 \e^{-\lambda x} \dd x \\ 
& \propto V(0)^2 \int_0^1  \cos^2\left(t \frac{\pi}{2} \right) m(\lambda) \dd t \ .
\end{align}
The asymptotic form of the initial condition stated in \eqref{eq_scaling_V0}, combined with the expression of $m(\lambda)$ given in \eqref{eq_prediction_moflambda}, yields finally the prediction
\begin{equation}
\mathbb{E}[\varrho] \underset{\lambda \to 4^-}{\propto} \e^{- \frac{2 \pi}{\sqrt{4 -\lambda}} } (4-\lambda)^{-3/2} \ .
\label{eq_rho_asymptotic}
\end{equation}
Note that all the derivatives of $\mathbb{E}[\varrho]$ vanish as $\lambda \to 4^-$ because of the essential singularity of the exponential term, the transition is thus of infinite order in the usual thermodynamic classification.

\begin{figure}
  \includegraphics[width=\columnwidth]{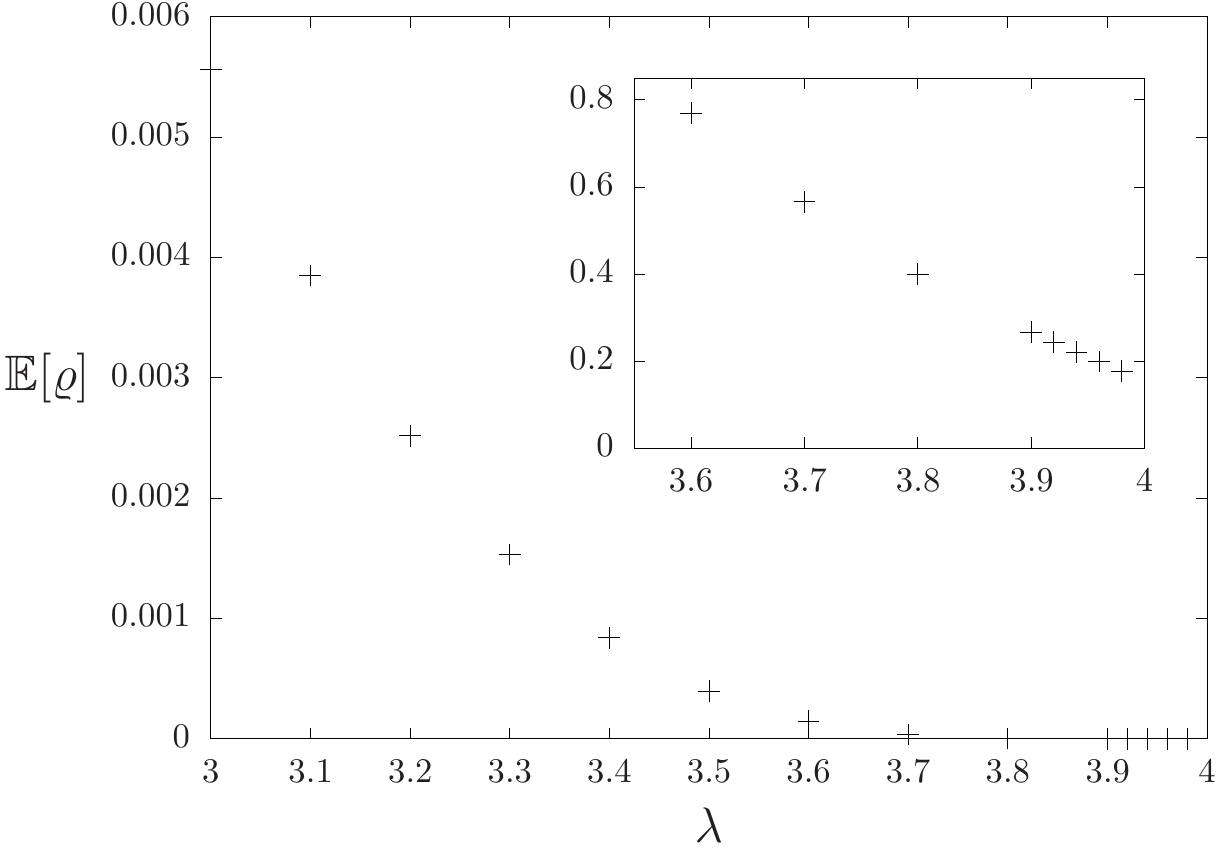}
\caption{The average reconstruction error $\mathbb{E}[\varrho]$ that vanishes continuously as $\lambda \to 4^-$. Inset: $\mathbb{E}[\varrho]\e^{ \frac{2 \pi}{\sqrt{4 -\lambda}} } (4-\lambda)^{3/2}$ as a function of $\lambda$, the convergence to a postive constant as $\lambda \to 4^-$ confirms \eqref{eq_rho_asymptotic}.
\label{fig_exp_rho}}
\end{figure}

We present in Fig.~\ref{fig_exp_rho} our numerical results for $\mathbb{E}[\varrho]$, computed by a numerical integration of the ODE and the integral expression in \eqref{eq_rho_exponential}. The main panel shows qualitatively that $\mathbb{E}[\varrho]$ is indeed very flat close to the transition; the rescaling performed in the inset is in agreement with the asymptotic form \eqref{eq_rho_asymptotic}.

\section{The symbol MAP case}
\label{sec:symbol_MAP}

We have discussed above the phase diagram of the problem, and distinguished in particular full and partial recovery phases, considering the block MAP estimator, i.e. the $\beta\to\infty$ version of the BP equations. The phases were thus defined according to whether the average reconstruction error $\mathbb{E}[\varrho_{\rm b}]$ vanished in the thermodynamic limit or not, the subscript b specifying the use of the block MAP estimator in the computation. However, we explained in Sec.~\ref{sec_def_inference} that the estimator that minimizes the average reconstruction error is the symbol MAP one, obtained with $\beta=1$, with an average reconstruction error denoted $\mathbb{E}[\varrho_{\rm s}]$. As $\mathbb{E}[\varrho_{\rm s}] \le \mathbb{E}[\varrho_{\rm b}]$ the phases shown to be of the full recovery type for $\beta\to\infty$ are certainly so also for the symbol MAP estimator, one can nevertheless wonder if the converse is true, namely if some choices of parameters yield $0=\mathbb{E}[\varrho_{\rm s}] < \mathbb{E}[\varrho_{\rm b}]$.

\begin{figure}
 \includegraphics[width=\columnwidth]{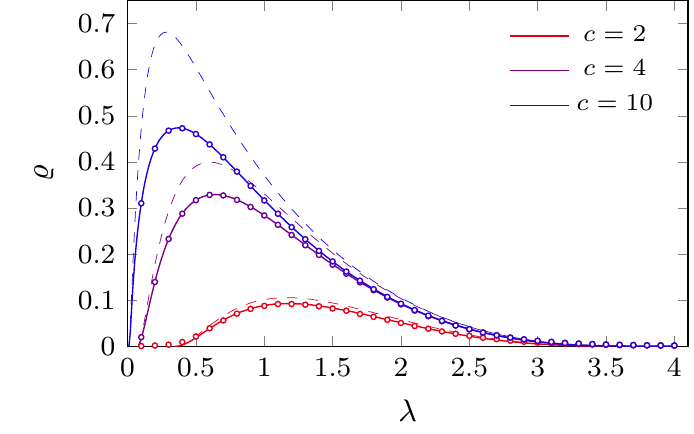}
\caption{The average reconstruction error for the symbol MAP estimator ($\beta=1$) with exponentially distributed planted weights, at different values of $c$. The solid lines have been obtained numerically solving the RDEs in Eq.~\eqref{rdebetaf} with $\beta=1$ using a population dynamics algorithm with $10^6$ fields. The dots corresponds to the same error rate estimated running BP over $10^4$ instances of the problem, on graphs with $N=10^3$ vertices. The dashed lines corresponds to the reconstruction error of the block MAP estimator presented in Fig.~\ref{fig:exp}, that are larger than the symbol MAP ones for the same value of $c$ (encoded by the color of the curve). \label{fig:bay}}
\end{figure}

We have investigated this question numerically, by solving with a population dynamics algorithm the RDEs \eqref{rdebetaf} with $\beta=1$, and computed $\mathbb{E}[\varrho_{\rm s}]$ from \eqref{eq_averho_2}. Our results are presented in Fig.~\ref{fig:bay}; for concreteness we have used the exponential distribution of Eq.~\eqref{expdis} for the planted weights, and several values of the average degree $c$ (the non-planted weight distribution being uniform on $[0,c]$). We found indeed that $\mathbb{E}[\varrho_{\rm s}] \le \mathbb{E}[\varrho_{\rm b}]$ (the block MAP results, previously presented on Fig.~\ref{fig:exp}, are drawn with dashed lines). Within our numerical accuracy the transition to the full recovery phases occur for the same values of the parameters in the symbol and block MAP cases; this is in agreement with a conjecture of~\cite{Moharrami2019}, see open question 1 therein.

\section{Future work}
\label{sec:conclusions}

Let us conclude by giving some thoughts on how our study could be extended. 
One could try to study the critical regime for generic distributions, i.e. extend the results of Sec.~\ref{sec:critical} that were obtained only for the exponential distribution and in the large degree limit. We expect the exponent $-1/2$ for the divergence of the median of the fields to be rather universal, but the form of the vanishing of $\mathbb{E}[\varrho]$ should be much more dependent on the details of the models. One motivation for this direction of research is the difficulty of an accurate numerical determination of the location of the phase transition, as discussed in Sec.~\ref{sec:resolution}. The numerical accuracy problems should be less stringent further away from $\bar\lambda$ inside the partial recovery phase, hence an extrapolation of $\mathbb{M}[\hat H]$, if one has a prediction for its functional form, should lead to more precise determinations of threshold parameters.

It would also be interesting to further investigate the possibility of discontinuous recovery phase transitions, for which the derivation presented in Sec.~\ref{sec:BRW} would fail. We did not find evidence for their occurence, but we cannot exclude this possibility because of the limited accuracy of our numerical results. Such situations might occur for contorted weight distributions, or if instead of Erd\H os-R\'enyi random graphs one hides the planted matching in a configuration model with some well-chosen degree distributions, for which~\cite{Bordenave2013} unveiled the existence of multiple BP fixed points.

The coincidence of the thresholds for full recovery of the symbol and block MAP estimators observed numerically in Sec.~\ref{sec:symbol_MAP} also calls for further investigation and for an analytical argument supporting (or disproving) it. This point is also connected to the apparent absence of statistical to computational gaps in this problem: the block MAP estimator, being a minimal weight perfect matching, can be determined in polynomial time~\cite{Edmonds1965}, and the results of~\cite{Bayati2008,Salez2009,Sanghavi2011,Bayati2011} strongly suggest that it can be asymptotically (in the large size limit) obtained by the $\beta \to \infty$ BP equations. An exact computation of the symbol MAP estimator is instead a computationally hard problem, but it is tempting to conjecture that the BP algorithm with $\beta=1$ reaches asymptotically the information theoretically optimal reconstruction error $\mathbb{E}[\varrho_{\rm s}]$.

A $k$-factor of a graph is a set of edges such that each node belongs to exactly $k$ edges of the factor; a perfect matching is thus a special case of this definition with $k=1$. It would therefore be interesting to study the planted $k$-factor problem for generic values of $k$. For $k=2$ the problem is related to the planted Hamiltonian cycle that was considered in~\cite{Bagaria2018}. The planted $k$-factor could also be studied using the cavity approach and the associated belief propagation equations. At variance with the matching case there is, for generic $k$, no efficient algorithm even for the block MAP estimator; this opens the possibility for computationally hard phases in such a generalization.

Another natural direction for future work is a rigorous proof of our results, notably of the threshold given in Eq.~\eqref{condizione_original} and the critical behaviour stated in Eq.~\eqref{eq_rho_asymptotic}. While the local-weak-convergence proof of \cite{Moharrami2019} can likely be extended to generic weights distribution and to the sparse graph settings, it is not clear how to control rigorously the solution of the recursive distributional equations, in particular the reasoning at the beginning of section \ref{sec:BRW}. The stochastic comparison argument explained in remark (iii) at the end of this section, and expanded upon in Appendix~\ref{appendix_ordering}, should provide a scheme for a rigorous proof of full recovery when $v>0$, the much more challenging question is to prove partial recovery when $v<0$. 

\subsection*{Acknowledgments}
We thank the authors of~\cite{Moharrami2019} for discussions and for sharing with us their results prior to publication. We also thank Florent Krzakala and Andrea Agazzi for useful discussions on the problem. This project has received funding from the European Union's Horizon 2020 research and innovation programme under the Marie Sk\l{}odowska-Curie grant agreement CoSP No 823748, and from the French Agence Nationale de la Recherche under grant ANR-17-CE23-0023-01 PAIL.

\appendix

\section{The reconstruction error for the block MAP estimator}
\label{app_rho_blockMAP}

We prove in this Appendix that the equality of the two expressions \eqref{eq_averho_2}  and \eqref{errore} of the average reconstruction error when $\beta\to\infty$ follows from the RDE \eqref{cavmapmatch}. We have thus to prove that $\mathbb{P}[\hat  H + \hat H' \le \hat \Omega  ] = \gamma \, \mathbb{P}[H + H ' > \Omega  ]$. We first notice that \eqref{cavmapmatch2} implies that for any real $x$ one has
$\mathbb{P}[H \ge x] =\mathbb{P}[\hat \Omega - \hat H \ge x] ( q + (1-q) \mathbb{P}[\hat H \ge x])$, hence
\begin{equation}
\mathbb{P}[\hat \Omega - \hat H \ge x] = \frac{\mathbb{P}[H \ge x] }{q + (1-q) \mathbb{P}[\hat H \ge x]} \ .
\end{equation}
Multiplying this expression by $-\frac{\dd}{\dd x} \mathbb{P}[\hat H \ge x]$, which is the density of the random variable $\hat H$, we obtain
\begin{align}
&\mathbb{P}[\hat  H + \hat H' \le \hat \Omega  ] = \int_{-\infty}^{+\infty} \left( -\frac{\dd}{\dd x} \mathbb{P}[\hat H \ge x]\right) \mathbb{P}[\hat \Omega - \hat H \ge x] \nonumber \\
&= \frac{1}{1-q} \int_{-\infty}^{+\infty} \left( -\frac{\dd}{\dd x} \ln( q + (1-q) \mathbb{P}[\hat H \ge x]) \right)\mathbb{P}[H \ge x] \nonumber \\
&=\frac{1}{1-q} \int_{-\infty}^{+\infty} \left( \frac{\dd}{\dd x} \mathbb{P}[ H \ge x]\right) \ln( q + (1-q) \mathbb{P}[\hat H \ge x]) \label{eq_rho_BM1}
\end{align}
where we performed an integration by part; the integrated term vanishes because there is no mass at infinity in the law of $H$ and $\hat H$.

We exploit now the other RDE \eqref{cavmapmatch1}, that gives
\begin{align}
\mathbb{P}[\hat H \ge x] & =\sum_{k=1}^\infty \frac{q}{1-q}\frac{[(1-q)\gamma]^k}{k!}  \mathbb{P}[\Omega- H \ge x ]^k \nonumber \\
& = \frac{q}{1-q} \left( \e^{(1-q) \gamma \mathbb{P}[\Omega- H \ge x ]} -1 \right) \ . \nonumber
\end{align} 
This yields
\begin{equation}
\ln( q + (1-q) \mathbb{P}[\hat H \ge x]) = - (1-q) \gamma \mathbb{P}[\Omega - H < x] \ ,
\label{eq_rho_BM2}
\end{equation}
where we used the equation $q=\e^{-\gamma (1-q)}$ to simplify the expression. Inserting \eqref{eq_rho_BM2} in \eqref{eq_rho_BM1} gives
\begin{align}
\mathbb{P}[\hat  H + \hat H' \le \hat \Omega  ] &= \gamma \int_{-\infty}^{+\infty} \left( -\frac{\dd}{\dd x} \mathbb{P}[ H > x]\right) \mathbb{P}[\Omega - H < x] \nonumber 
\\ &= \gamma \, \mathbb{P}[H + H ' > \Omega  ] \ , \nonumber
\end{align}
which proves our claim. Note that this derivation relies crucially on the hypothesis that $H$ and $\hat{H}$ have a continuous distribution, which allowed to introduce their density and to perform integration by parts to connect the two terms of \eqref{eq_averho_2}. We expect this to be the case when the effective weight distribution is continuous, in such a way that the minimal weight perfect matching is unique (on a finite graph); a counterexample is discussed in Sec.~\ref{sec:powerlaw}.

\section{On the simplified RDE in Sec.~\ref{sec:BRW}}\label{app:rde}

We provide in this Appendix some additional details about the simplified RDE defined in Sec.~\ref{sec:BRW}; we first give an heuristic justification of the velocity \eqref{vgen} of the leftmost particle of a BRW, then we detail the stochastic ordering argument that leads to the divergence of $\hat H^{(n)}$ when $v>0$.

\subsection{Heuristic derivation of the velocity in the BRW process}

We will present a reasoning typical of the physics literature on front propagation in reaction-diffusion systems and equations of the FKPP type, see for instance~\cite{Brunet1997,Majumdar2000,Ebert2000}, that leads to the expression \eqref{vgen} for the velocity of the leftmost particle of the BRW. 

We define the cumulative distribution function of $\hat K^{(n)}$ as $F(x,n)=\mathbb{P}[\hat K^{(n)}\le x]$. For a given time $n$ this is an increasing function of $x$, from $0$ to $1$ as $x$ increases from $-\infty$ to $+\infty$. The RDE \eqref{simpitera} translates into an evolution equation for $F$ as the discrete time increases,
\begin{equation}
F(x,n+1) = 1- \sum_{k=1}^\infty \pi_k \left(1-\int F(x-\Xi,n) \chi(\Xi) \dd \Xi \right)^k \ ,
\nonumber
\end{equation}
where $\pi_k$ is the probability law of the random variable $Z$, and $\chi$ the density of $\Xi$. We assume that at large times $F$ exhibits a front propagating at a velocity $v$, and denote $F_v$ the shape of the front in the reference frame moving at this velocity: $F(z+vn,n) \to F_v(z)$ as $n \to \infty$. This gives the following equation on $F_v$:
\begin{equation}
F_v(z-v) = 1- \sum_{k=1}^\infty \pi_k \left(1-\int F_v(z-\Xi) \chi(\Xi) \dd \Xi \right)^k \ ,
\nonumber
\end{equation}
which is equivalent to the RDE \eqref{eq_BRW_Y} on the limit random variable $L$. When $z \to - \infty$ the distribution function vanishes, in this limit we can thus linearize the equation on $F_v$, which yields:
\begin{equation}
F_v(z-v) = \left(\sum_{k=1}^\infty \pi_k k  \right) \int F_v(z-\Xi) \chi(\Xi) \dd \Xi  \ .
\end{equation}
This linear (integral) equation admits solutions of the form $F_v(z)=\e^{\theta z}$, with $\theta >0$ to respect the increasing character of distribution functions, if $\theta$ and $v$ obey the condition
\begin{equation}
\e^{-\theta v} = \left(\sum_{k=1}^\infty \pi_k k  \right) \int \e^{-\theta \Xi} \chi(\Xi) \dd \Xi  \ ,
\end{equation}
which gives a relation $v=v(\theta)$ corresponding to \eqref{vgen}. The linearized equation thus admits a family of solutions parametrized by the tail exponent $\theta >0$, corresponding to velocities $v(\theta)$. The delicate point in this reasoning, for which we refer the reader to the literature, is the justification of the minimum velocity selection principle, namely the fact that the relevant solution of the full non-linear equation on $F_v$ is the one minimizing $v(\theta)$, as stated in \eqref{vgen}.

Note that the minimizer $\theta_*$ of $v(\theta)$ corresponds to a double root of the characteristic equation of the linearized equation on $F_v$, which thus admits as solutions the linear combinations of $\e^{\theta_* z}$ and $z \, \e^{\theta_* z}$. This enlightens the statement made in \eqref{eq_left_tail_Y} for the
left tail behavior of the limit random variable $L$, obtained rigorously in~\cite{Aidekon2013,Bramson2016}.

\subsection{Stochastic ordering argument}
\label{appendix_ordering}
Let us prove here the claim made in remark (iii) of Sec.~\ref{sec:BRW}, namely that the sequence of random variables $\hat K^{(n)}$ defined in \eqref{simpitera} by the simplified RDE provides a stochastic lower-bound for the sequence $\hat H^{(n)}$ of the complete RDE \eqref{eq_RDE_iterations}. We recall that a random variable $X$ is said to be stochastically smaller than a random variable $Y$ if and only if $\mathbb{P}[X>x] \le \mathbb{P}[Y>x] $ for all $x$, which we denote $X \preceq Y$. A very useful equivalent characterization of this property~\cite{Lindvall2012} is the existence of a coupling $(\hat X,\hat Y)$, i.e. a random vector with marginal laws equal to those of $X$ and $Y$ respectively, such that $\mathbb{P}[\hat{X} \le \hat Y]=1$. In other words if $X \preceq Y$ one can consider that $X$ and $Y$ are defined on the same probability space and that $X \le Y$ with probability one.

To simplify the comparison between \eqref{simpitera} and \eqref{eq_RDE_iterations} we break the iteration \eqref{simpitera} in two steps and define another sequence of random variables $K^{(n)}$, with
\begin{subequations} 
\thinmuskip=0mu
 \begin{align}
  \hat K^{(n+1)}&\stackrel{\mathrm{d}}{=}\min_{1\leq i\leq Z}\left[\Omega_{i}-K_{i}^{(n)}\right]\ ,  \label{eq_hK_np1} \\
  K^{(n)}&\stackrel{\mathrm{d}}{=} \hat \Omega-\hat K^{(n)} \label{eq_Kn} \ .
 \end{align}
\end{subequations}

We claim that if the initial condition for $\hat H^{(n)}$ and $\hat K^{(n)}$ is the same, namely $\hat H^{(0)}=\hat K^{(0)}=0$, then for all $n \ge 0$ one has $\hat K^{(n)} \preceq \hat H^{(n)} $ and $H^{(n)} \preceq K^{(n)} $. The proof is obtained by two induction steps on $n$. 

Suppose that $H^{(n)} \preceq K^{(n)}$; then one can couple \eqref{eq_RDE_iterations_a} and \eqref{eq_hK_np1} by taking the same random variables $Z$ and $\Omega_i$ in both, and by using the existence of the coupling of $H^{(n)}$ and $K^{(n)}$ to ensure that $H_i^{(n)} \le K_i^{(n)}$ with probability one, for all $i \in \{1,\dots,Z\}$. This yields a coupling of $\hat H^{(n+1)}$ and $\hat K^{(n+1)}$ such that $\hat K^{(n+1)} \le \hat H^{(n+1)}$ with probability one, which proves that $\hat K^{(n+1)} \preceq \hat H^{(n+1)}$.

Assume now that $\hat K^{(n)} \preceq \hat H^{(n)} $, and couple \eqref{eq_RDE_iterations_b} and \eqref{eq_Kn} by taking the same random variable $\hat \Omega$ in both, the same $\hat H_1^{(n)} $ in the two alternatives of \eqref{eq_RDE_iterations_b}, and by ensuring that $\hat K^{(n)} \le \hat H_1^{(n)} $ with probability one. We have thus coupled $H^{(n)}$ and $K^{(n)}$ in such a way that $H^{(n)} \le K^{(n)}$ with probability one, hence yielding $H^{(n)} \preceq K^{(n)}$.

As the initial condition obviously satisfies the induction hypothesis $\hat K^{(0)} \preceq \hat H^{(0)} $ this proves our claim: for all $n$ one has $\hat K^{(n)} \preceq \hat H^{(n)} $, and in particular if $v>0$ the divergence to $+ \infty$ of the sequence $\hat K^{(n)} $ as $n \to \infty$ implies the one of $\hat H^{(n)} $.

\bibliography{bibliografia.bib}
\end{document}